\documentclass[apj]{emulateapj}

\shorttitle{3D MHD Simulations of Winds of Solar-Like Stars}
\shortauthors{Vidotto et al.}

\begin{document}

\title{Three-dimensional Numerical Simulations of Magnetized Winds of Solar-Like Stars}

\author{A. A. Vidotto}
\affil{University of S\~ao Paulo, Rua do Mat\~ao 1226, S\~ao Paulo, SP, Brazil, 05508-090} 
\affil{George Mason University, 4400 University Drive, Fairfax, VA, USA, 22030-4444}
\email{aline@astro.iag.usp.br}              

\author{M. Opher}
\affil{George Mason University, 4400 University Drive, Fairfax, VA, USA, 22030-4444}

\author{V. Jatenco-Pereira}
\affil{University of S\~ao Paulo, Rua do Mat\~ao 1226, S\~ao Paulo, SP, Brazil, 05508-090}

\and

\author{T. I. Gombosi}
\affil{University of Michigan, 1517 Space Research Building, Ann Arbor, MI, USA, 48109-2143}

\begin{abstract}
By means of self-consistent three-dimensional (3D) magnetohydrodynamics (MHD) numerical simulations, we analyze magnetized solar-like stellar winds and their dependence on the plasma-$\beta$ parameter (the ratio between thermal and magnetic energy densities). This is the first study to perform such analysis solving the fully ideal 3D MHD equations. We adopt in our simulations a heating parameter described by $\gamma$, which is responsible for the thermal acceleration of the wind. We analyze winds with polar magnetic field intensities ranging from $1$ to $20$~G. We show that the wind structure presents characteristics that are similar to the solar coronal wind. The steady-state magnetic field topology for all cases is similar, presenting a configuration of {\it helmet streamer}-type, with zones of closed field lines and open field lines coexisting. Higher magnetic field intensities lead to faster and hotter winds. For the maximum magnetic intensity simulated of $20$~G and solar coronal base density, the wind velocity reaches values of $\sim 1000$ km~s$^{-1}$ at $r \sim 20~r_0$ and a maximum temperature of $\sim 6 \times 10^6$~K at $r\sim 6~r_0$. The increase of the field intensity generates a larger ``dead zone'' in the wind, i. e., the closed loops that inhibit matter to escape from latitudes lower than $\sim 45^o$ extend farther away from the star. The Lorentz force leads naturally to a latitude-dependent wind. We show that by increasing the density and maintaining $B_0=20$~G, the system recover back to slower and cooler winds. For a fixed $\gamma$, we show that the key parameter in determining the wind velocity profile is the $\beta$-parameter at the coronal base. Therefore, there is a group of magnetized flows that would present the same terminal velocity despite of its thermal and magnetic energy densities, as long as the plasma-$\beta$ parameter is the same. This degeneracy, however, can be removed if we compare other physical parameters of the wind, such as the mass-loss rate. We analyze the influence of $\gamma$ in our results and we show that it is also important in determining the wind structure.
\end{abstract}


\section{INTRODUCTION}
Studies of the solar corona (SC) have played a crucial role in understanding stellar winds in general. Due to our privileged position immersed in the solar wind, researchers have had access to a great quantity of data that allow a detailed understanding of the physics that is operating in the Sun. Recent sophisticated observations, e. g., via SOHO and Ulysses \citep[among others]{1995JGR...10019893M, 1996GeoRL..23.3267S, 1998GeoRL..25.3109J, 2006A&A...455..697W}, have shown that the SC is a highly complex system. It consists of long-lived features, like the fast and slow solar wind, streamers and coronal holes, and also of short-lived features, like the coronal mass ejections, solar flares and sun-spots. Hence, if one could separate the long-lived structures from the short-lived ones, one could, in principle, come to a better understanding of the SC itself and other similar outflows. In the present paper, we aim to investigate in detail the behavior of the long-lived coronal features in solar-like stellar winds. 

Direct measurements of tenuous coronal winds for other stars rather than the Sun have proved to be very difficult to do, although indirect detections of stellar coronal winds have been performed \citep{2005ApJ...628L.143W}. As it occurs in the SC, it is very probable that the magnetic field is playing an important role in coronal winds of solar-like stars. Magnetic activity has not only been detected in other stars \citep{2002A&A...389..191B, 2006ApJ...644..497R, 2006ApJ...646L..73P, 2007MNRAS.376.1145W}, but also in stars similar to the Sun \citep{1980ApJ...236L.155R, 2005MNRAS.361..837P}.

In the absence of a magnetic field, a non-rotating corona expands spherically \citep{1958ApJ...128..664P, 1994ApJ...432L..55V}. A different picture, however, is expected if the magnetic energy density at the base of the corona is at {\it least} of the same amount of the thermal energy density. 

A particular case of equal magnetic and thermal energy densities was explored by Pneuman \& Kopp (1971, \citet{1971SoPh...18..258P} from now on) for the case of the magnetized solar wind. By means of a numerical iterative method, they concluded that the SC is composed of two different magnetic structures in large scale: a region of closed magnetic field loops near the star at low latitudes and at high latitudes, open field lines that cannot restrain the expanding gas. 

Astrophysical outflows have long been studied \citep{1958ApJ...128..664P, 1967ApJ...148..217W, 1968MNRAS.138..359M, 1975ApJ...196..837N}. Several analytical studies were made toward the understanding of an expanding magnetized corona. For instance, \citet{1986ApJ...302..163L} found a class of radial analytic solutions for magnetized steady-state winds by removing the latitude dependence of the problem. By varying the global magnetic field geometry and other parameters of their model, such as the velocity at the base of the wind, they analyzed the behavior of flows along open magnetic field lines. \citet{1991A&A...249..156T} relaxed the assumption of a polytropic equation of state, but neglected the meridional component of the flow. \citet{1994A&A...287..893S} found a new class of analytical solutions calculating the exact shape of a field line along the flow rather than assuming a global magnetic field geometry. They showed that for a rotating young stellar object the shape of the field lines is an important parameter for the formation of a collimated jet/non-collimated wind. In a more recent work, \citet{2001A&A...371..240L} constructed exact solutions for a rotating, magnetized wind. They verified that the wind is highly non-spherically symmetric. However, as they pointed out, their model has limitations, as, for instance, meridional flows and the meridional component of the magnetic field were neglected.  

Meanwhile, numerical studies were carried out with increasing level of sophistication. \citet{1993MNRAS.262..936W} studied the influence of the stellar rotation on the wind structure and acceleration performing 2D simulations. Solar wind parameters were used, except for the uniform angular velocity at the surface of the star that was varied as to analyze the effects of the centrifugal force. They showed that if the angular velocity is increased more than ten times the solar value, the centrifugal force becomes comparable to the thermal force, influencing the meridional structure of the wind. \citet{1999A&A...343..251K} presented detailed one-dimensional (1D) and two-dimensional (2D) ideal MHD numerical simulations of a polytropic, axisymmetric wind. In a posterior work, \citet{2000ApJ...530.1036K} modeled stellar axisymmetric rotating outflows by solving the ideal MHD equations, investigating the effects of open and closed magnetic field lines in the wind. By varying the extension of the closed field line region, as well as the intensity of the magnetic field, they showed that the global wind structure is modified. \citet{2003PhDT.........1U}, working towards the modeling of hot-star winds, presented a simulation of the solar wind using time-dependent, axisymmetric, MHD simulation. He showed that the MHD modeling was consistent with the work of \citet{1971SoPh...18..258P}. Modeling the propagation of a coronal mass ejection from the inner solar corona to 1 AU, \citet{2000JGR...10525053G} numerically reproduced the steady-state bi-modal nature of the solar wind with a prescribed {\it ad-hoc} heating mechanism. Working on the modeling of the solar wind, \citet{2003ApJ...595L..57R} simulated the 3D structure of the solar wind under steady-state conditions, using solar magnetogram data as input parameters for the initiation of the wind, and considering a variable heating mechanism. \citet{2007ApJ...654L.163C} extended  \citet{2003ApJ...595L..57R}'s work considering, as a heating mechanism, a radial dependence of the ratio of specific heats, $\gamma = \gamma (r)$, as to reproduce the observed bi-modality of the velocity of the solar wind. 

Despite all the notable evolution of both analytical and numerical studies performed in the last decades, we are far from a satisfactory 3D MHD description of a magnetized wind. Several approximations were made in order to make the system analytically and numerically tractable (e. g., neglecting meridional flows, assuming a polytropic equation of state, assuming a magnetic field topology). 

In the present study, we investigate the influence of the magnetic field in solar-like stellar winds with different plasma-$\beta$ (the ratio between thermal and magnetic energy densities). We solve the fully 3D MHD equations with the temporal evolution of the energy equation. Therefore, the topology of the field is not restricted and the steady-state arises from the dynamical interplay of the outflow and the field. Also, meridional flows arise naturally in the system. We neglect the stellar rotation. The results presented here are thus valid for non-rotators or slow rotators in the region where the toroidal component of the field is still much smaller than the poloidal component.

The paper is organized as follows. In \S 2, we present the numerical scheme used and in \S 3, the results obtained. \S 4 is dedicated to conclusions and discussion.

\section{THE NUMERICAL MODEL} 
To perform the simulations, we make use of the Block Adaptive Tree Solar-wind Roe Upwind Scheme (BATS-R-US), a three-dimensional MHD numerical code developed at the Center for Space Environment Modeling at University of Michigan \citep{1999JCoPh.154..284P}.

BATS-R-US uses a computational domain that is block-based, consisting of Cartesian blocks of cells that can be adaptively refined. It has been widely used to simulate the Earth's magnetosphere \citep{2006AdSpR..38..263R}, the heliosphere \citep{2003ApJ...595L..57R, 2007ApJ...654L.163C}, the outer-heliosphere \citep{1998JGR...103.1889L, 2003ApJ...591L..61O, 2004ApJ...611..575O}, coronal mass ejections \citep{2004JGRA..10901102M,2005ApJ...627.1019L}, and the magnetosphere of planets \citep{2004JGRA..10911210T,2005GeoRL..3220S06H}, among others. In this work, we adapted the version for the outer heliosphere \citep{2003ApJ...591L..61O, 2004ApJ...611..575O} to study the problem of the wind of a solar-like star. 

The code solves the ideal MHD equations, that in the conservative form are given by
\begin{equation}
\label{eq:continuity_conserve}
\frac{\partial \rho}{\partial t} + \nabla\cdot \left(\rho {\bf u}\right) = 0
\end{equation}
\begin{equation}
\label{eq:momentum_conserve}
\frac{\partial \left(\rho {\bf u}\right)}{\partial t} + \nabla\cdot\left[ \rho{\bf u\,u}+ \left(p + \frac{B^2}{8\pi}\right)I - \frac{{\bf B\,B}}{4\pi}\right] = \rho {\bf g}
\end{equation}
\begin{equation}
\label{eq:bfield_conserve}
\frac{\partial {\bf B}}{\partial t} + \nabla\cdot\left({\bf u\,B} - {\bf B\,u}\right) = 0
\end{equation}
\begin{equation}
\label{eq:energy_conserve}
\frac{\partial\varepsilon}{\partial t} +  \nabla \cdot \left[ {\bf u} \left( \varepsilon + p + \frac{B^2}{8\pi} \right) - \frac{\left({\bf u}\cdot{\bf B}\right) {\bf B}}{4\pi}\right] = \rho {\bf g}\cdot {\bf u} \, ,
\end{equation}
where $\rho$ is the mass density, ${\bf u}$ the plasma velocity, ${\bf B}$ the magnetic field, $p$ the gas pressure, ${\bf g}$ the gravitational acceleration due to the central body, and  $\varepsilon$ is the total energy density given by 
\begin{equation}
\varepsilon=\frac{\rho u^2}{2}+\frac{p}{\gamma-1}+\frac{B^2}{8\pi} \, .
\end{equation}
We consider ideal gas, so $p=\rho k_B T/(\mu m_p)$, where $k_B$ is the Boltzmann constant, $T$ is the temperature, $\mu m_p$ is the mean mass of the particle, and $\gamma$ is the ratio of the specific heats.

Due to the lack of knowledge of all the detailed processes that take place in a stellar wind, it is difficult to estimate all the mechanisms that modify the heat content of the wind (e. g., conduction, radiation, mechanical dissipation of energy that is transferred to the plasma). MHD waves and turbulence are known to play an important role on solar wind acceleration and heating \citep{1999ApJ...523L..93M, 2005ApJS..156..265C, 2007ApJS..171..520C}. For this reason, they are often included in stellar wind models as well \citep[e.g., ][]{2000ApJ...528..965A, 2006ApJ...639..416V, 2007ApJ...659.1592S}. The solar-stellar wind connection is usually made by scaling the observed solar wind characteristics (e.g., the wave flux and spectral slope, solar magnetograms) to other stars \citep{1989A&A...209..327J, 2001ApJ...547..475S}. However, it is not clear how these characteristics scale to other solar-like stars, in this first analysis we start with a simpler treatment of the mechanisms of acceleration of the wind. We adopt, therefore, an approach similar to \citet{2003ApJ...595L..57R} who considered that $\gamma$ is associated with ``turbulent'' internal degrees of freedom, in a way analogous to the Sun, where a significant amount of energy is stored in the form of waves and turbulent fluctuations.

\subsection{The Grid Adopted}
\begin{figure*}
  \includegraphics[scale=0.26]{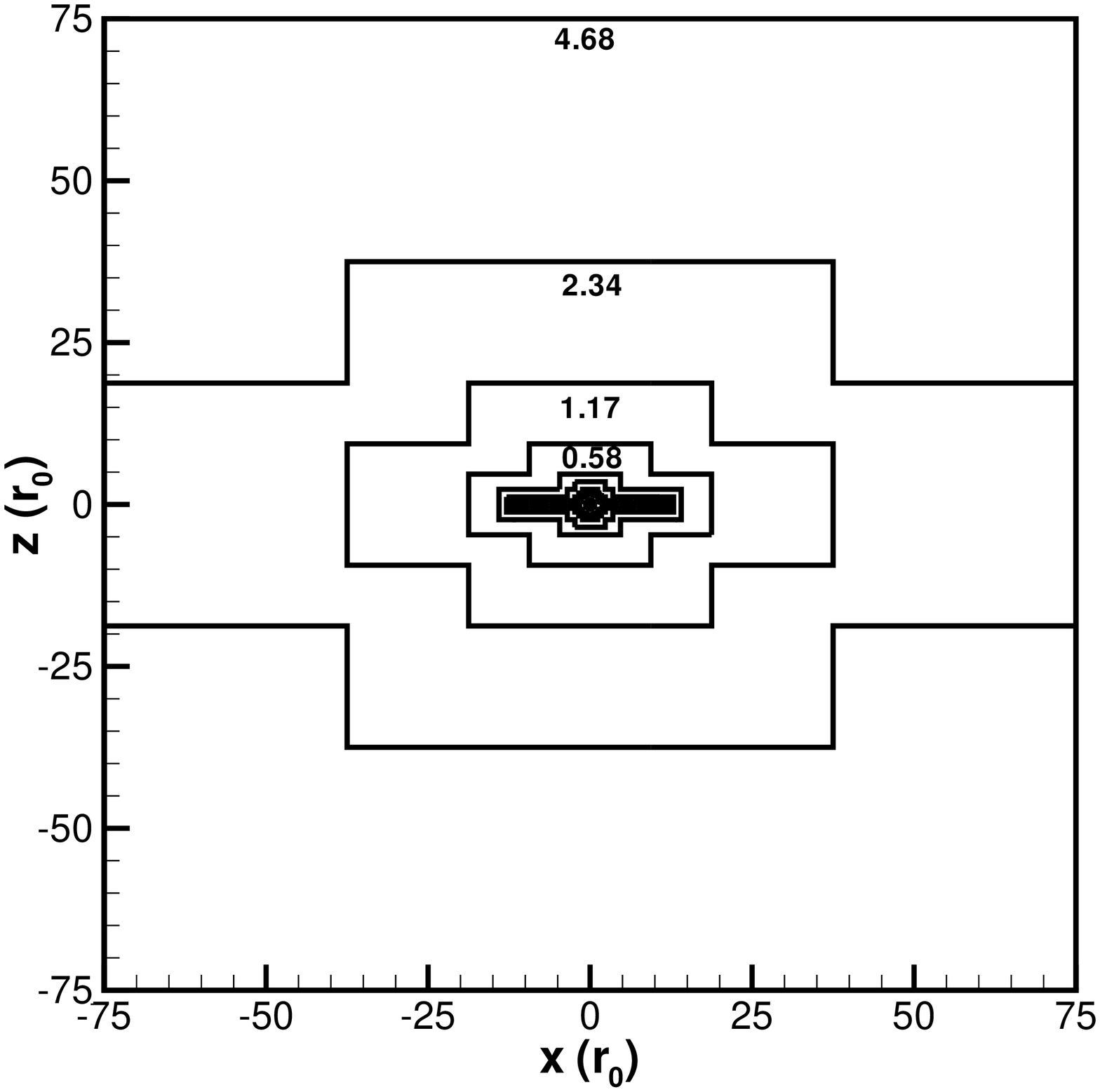}%
  \includegraphics[scale=0.26]{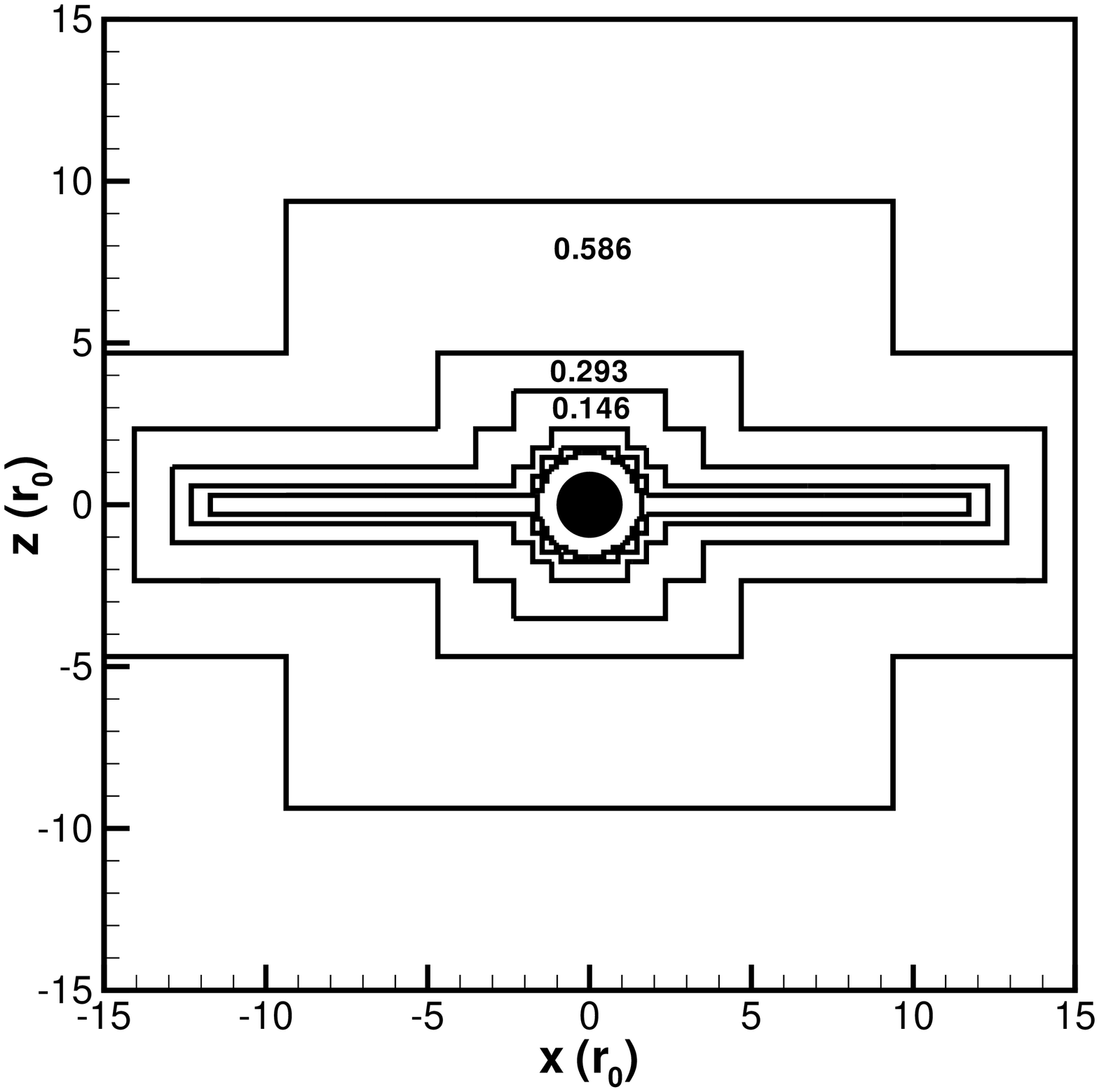}%
  \includegraphics[scale=0.26]{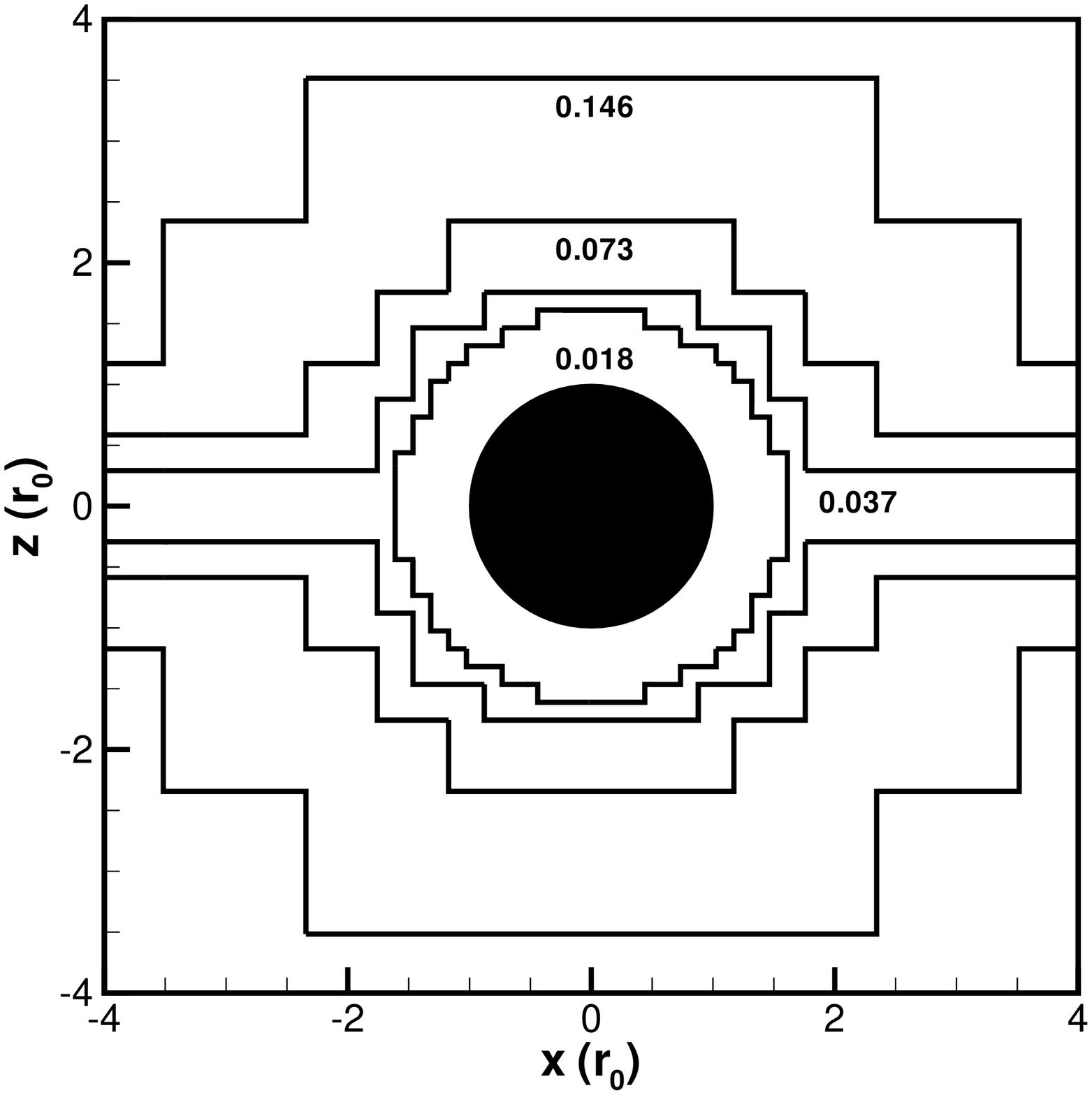}%
  \caption{Three different views of the grid. Inside each region, we indicate the maximum cell size of the region (in $r_0$ units). \label{grid} }
\end{figure*}

For all the simulations done, we adopted the same grid resolution. Initially, the simulation domain is refined in five levels. Other five refinement levels that are body-focused and focused on the equatorial plane (current sheet region) are applied next. Finally, an additional level is applied to the body. There are $9.1 \times 10^6$ cells in the domain. The smallest cell size is $0.018~r_0$, located around the central body. The maximum cell size is $4.68~r_0$. The cell size near the current sheet is $0.036~r_0$. The grid is Cartesian $\{ x,y,z \}$ and the center of the star is placed at the origin. The axes $x$, $y$ and $z$ extend from $-75~r_0$ to $75~r_0$. The grid can be seen in Fig.~\ref{grid}. 

\subsection{The Initial Conditions}
The simulation domain is initialized with a solar-like star at its center with $1~M_\odot$ and $r_0=1~R_\odot$. The grid is initialized with a 1D hydrodynamical (HD) wind for a totally ionized plasma of hydrogen. he initial temperature profile is set according to the value of $\gamma$ adopted. This solution is dependent solely on the choice of the base temperature of the wind and the only physical possible solution is the one that becomes supersonic when passing through the critical radius \citep{1958ApJ...128..664P}. Due to conservation of mass of a steady wind, we obtain the density profile from the radial velocity profile $u_r (r)$. Initially, we chose a coronal base temperature of $T_0 = 1.56 \times 10^6$~K and a base density of $\rho_0 = 1.544 \times 10^{-16} $~g~cm$^{-3}$ (we vary $\rho_0$ in the second and third sets of simulations presented later on). 

The simulations are initialized with a bipolar magnetic field configuration described in spherical coordinates $\{ r, \theta, \varphi \}$ by
\begin{equation}
\label{eq:dipole}
{\bf B} =  \frac{B_0 r_0 ^3}{r ^3} \left(\cos \theta , \frac12 \sin \theta, 0 \right) \, ,
\end{equation}
where $B_0$ is the magnetic field intensity at the poles, $r$ is the radial coordinate, $\theta$ is the co-latitude, and $\varphi$ is the azimuthal angle measured in the equatorial plane. The system is then  evolved in time until steady-state is achieved.

\subsection{The Boundary Conditions}
The inner boundary of the system is considered to be the base of the wind at $r=r_0$ and its conditions are dependent on local flow conditions: plasma can freely leave the reservoir (i.e., the base of the coronal wind), but no ``backflow" is allowed. Fixed boundary conditions were adopted at $r=r_0$. The outer boundary has outflow conditions, i. e., a zero gradient is set to all the primary variables.

\section{RESULTS}
Table~\ref{table:1} presents the parameters used for the simulations performed. S00 is a purely HD simulation included for comparison purposes. The simulations are divided in three sets. The first set of simulations, composed by simulations S01 to S05, aims to investigate the effect of the magnetic energy density on the wind. We increased $B_0$ from $1$ to $20$~G while maintaining the other initial values fixed: $\beta_0$, the ratio between the surface thermal pressure $p_0$ and the magnetic pressure evaluated at the pole
\begin{equation}
\label{eq:beta0}
\beta_0 = \frac{8 \pi p_0}{{B_0} ^2} \, ,
\end{equation}
is decreased  from $1$ to $1/400$. S01 is similar to the case studied by \citet{1971SoPh...18..258P} with $\beta_0 =1$. The initial HD wind is the same for all the simulations of the first set ($\rho_0=\rho_c=1.544\times 10^{-16}$~g~cm$^{-3}$, $T_0=1.56 \times 10^{6}$~K, as can be seen in Table~\ref{table:1}). In the second set of simulations, from simulations S06 to S09, $B_0$ is fixed at $20$~G and we vary $\rho_0$ from $1$ to $400$ times the solar density, $\rho_c= 1.544 \times 10^{-16}$~g~cm$^{-3}$, thus increasing $\beta_0$ back to $1$ in S09. In both sets of simulations, we adopted the value $\gamma = 1.01$. We then consider a third set of simulations, where parameters similar to the first and second sets are used, but we adopt $\gamma=1.1$.

\begin{deluxetable}{c c c c c c }  
\tablewidth{0pt}
\tablecaption{The complete set of simulations  \label{table:1}}  
\tablehead{\colhead{Name} & \colhead{$T_0$ (MK)} & \colhead{$\rho_0/\rho_c$\tablenotemark{*}} & \colhead{$B_0$ (G)} & \colhead{$\beta_0$} & \colhead{$\gamma$}} 
\startdata
S00 & $1.56$ & $1$& $0 $  & $\infty$ & $1.01$\\
\cutinhead{First Set of Simulations}
S01 & $1.56$ & $1$& $1 $  & $1    $ & $1.01$\\
S02 & $1.56$ & $1$& $5 $  & $1/25 $ & $1.01$\\
S03 & $1.56$ & $1$& $10$  & $1/100$ & $1.01$\\
S04 & $1.56$ & $1$& $15$  & $1/225$ & $1.01$\\
S05 & $1.56$ & $1$& $20$  & $1/400$ & $1.01$\\ 
\cutinhead{Second Set of Simulations}
S06 & $1.56$ & $1.8$ &$ 20$  &$1/225  $ & $1.01$\\
S07 & $1.56$ & $4$ &$ 20$  &$1/100   $ & $1.01$\\
S08 & $1.56$ & $16$ &$ 20$  &$1/25     $ & $1.01$\\
S09 & $1.56$ & $400$ &$ 20$  &$1$ & $1.01$\\
\cutinhead{Third Set of Simulations}
S01b & $1.56$ & $1$  & $1$  & $1$ & $1.1$ \\
S02b & $1.56$ & $1$  & $5 $  & $1/25 $ & $1.1$ \\
S08c & $1.56$ & $16$  & $20$  & $1/25 $ & $1.1$ \\
S09c & $1.56$ & $400$ & $ 20$ &$1$ & $1.1$ 
\enddata
\tablenotetext{*}{$\rho_c = 1.544 \times 10^{-16}$~g~cm$^{-3}$}
\end{deluxetable}

\subsection{The First Set of Simulations}

\subsubsection{Evolution to Steady-State}
The Lorentz force ${\bf F_{B}}$ acting on the flow is given by the difference between the magnetic tension, ${\bf \mathcal{T}_B}$, and a non-isotropic gradient of the magnetic pressure $P_B$
\begin{equation} \label{eq.lorentz}
{\bf F_{B}} = {\bf \mathcal{T}_B} - {\bf \nabla_\perp} P_B  \, ,
\end{equation}
where ${\bf \nabla_\perp}$ is the component of the gradient perpendicular to the magnetic field ${\bf B}$. As the magnetic field lines are more curved at low latitudes, the tension becomes more important as one approaches the equatorial plane. Hence, opening a bipolar field structure at low latitudes, where lines are more curved, is more difficult than opening a bipolar field topology at high latitudes. As both ${\bf \mathcal{T}_B}$ and ${\bf \nabla_\perp} P_B$ are latitude-dependent, so is ${\bf F_B}$. We therefore expect from the interaction of the outflow and the magnetic field a latitude-dependent wind.

\begin{figure*}
\includegraphics[scale=0.30]{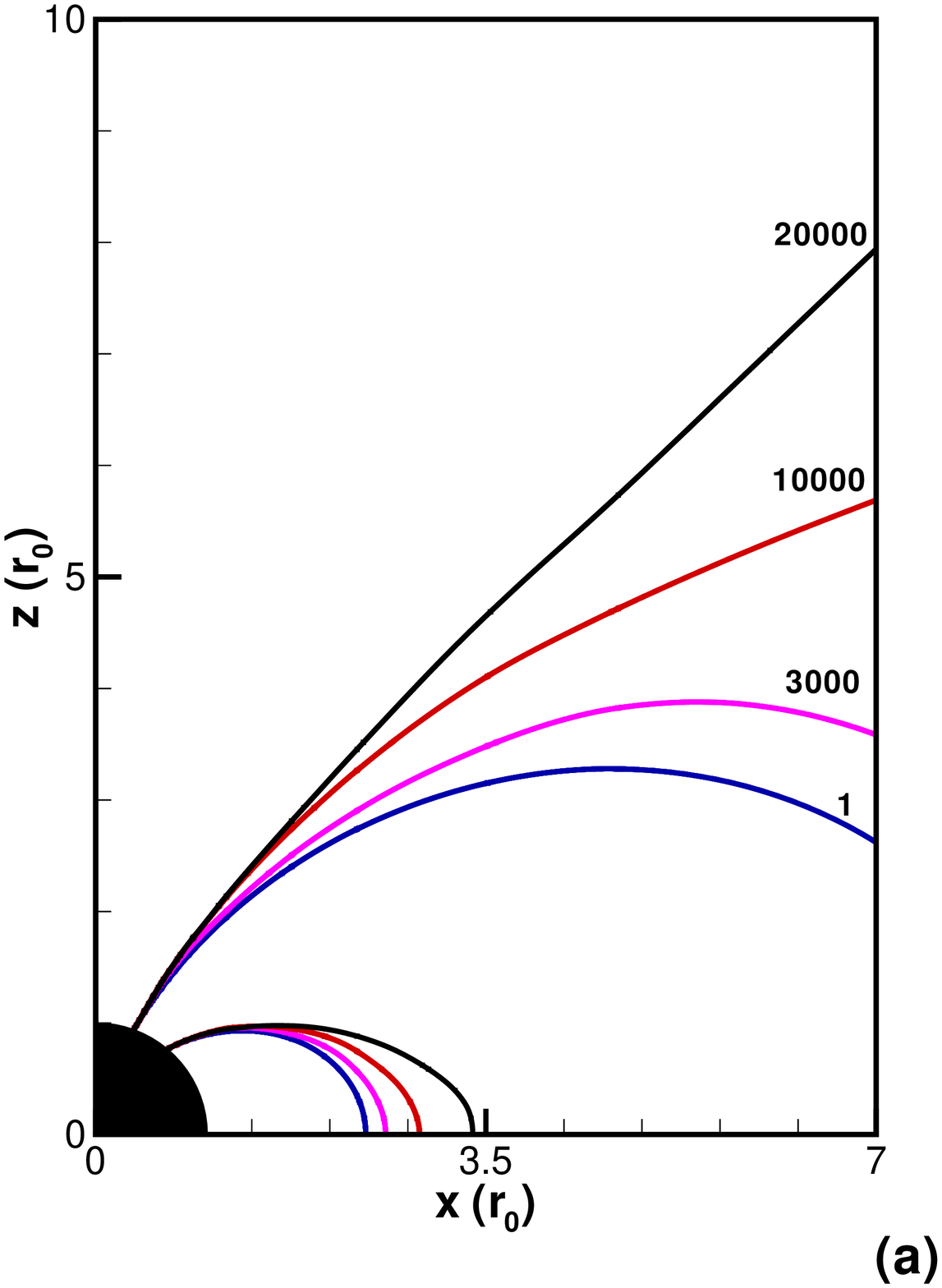}%
\includegraphics[scale=0.30]{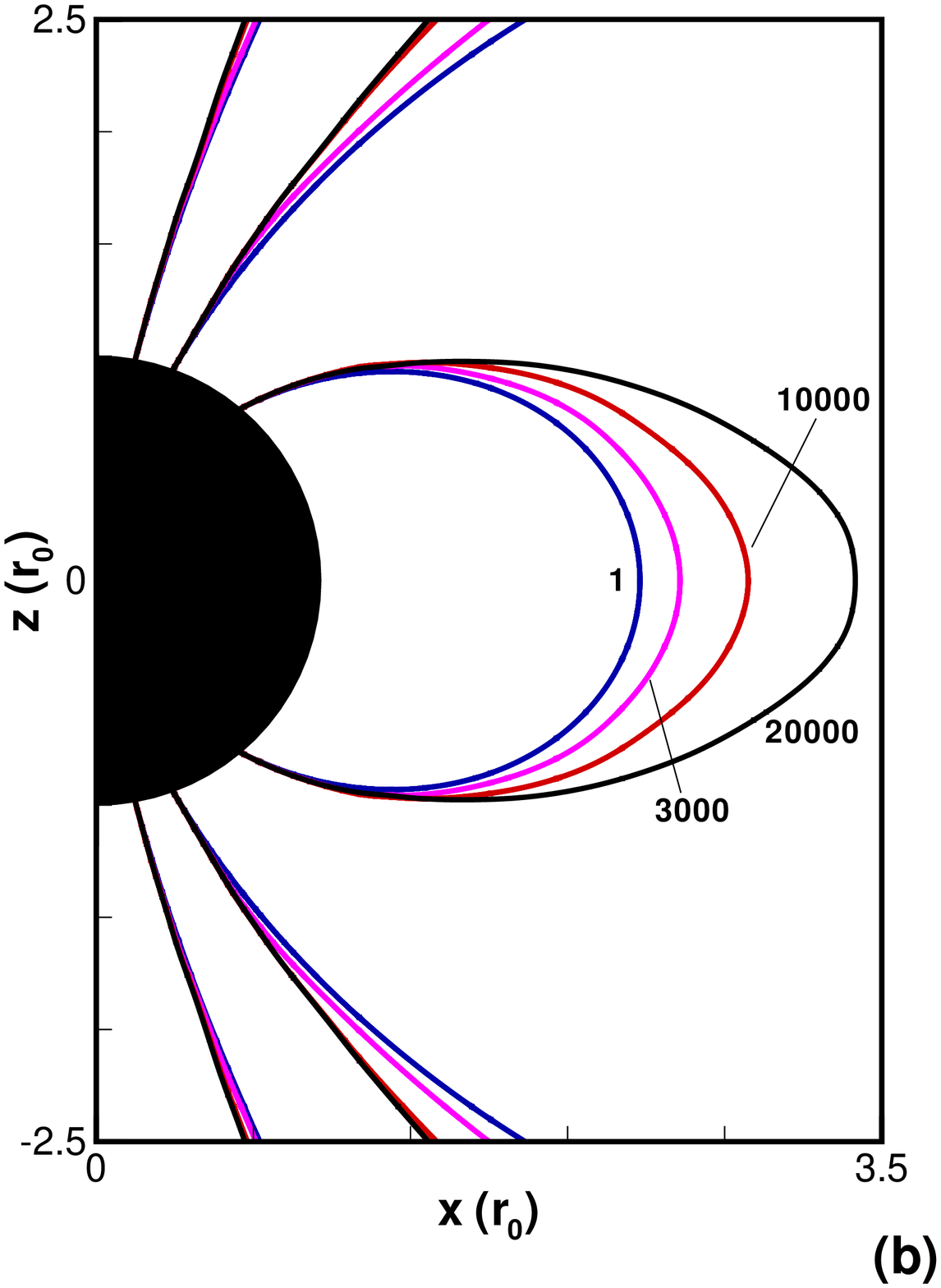}%
\includegraphics[scale=0.30]{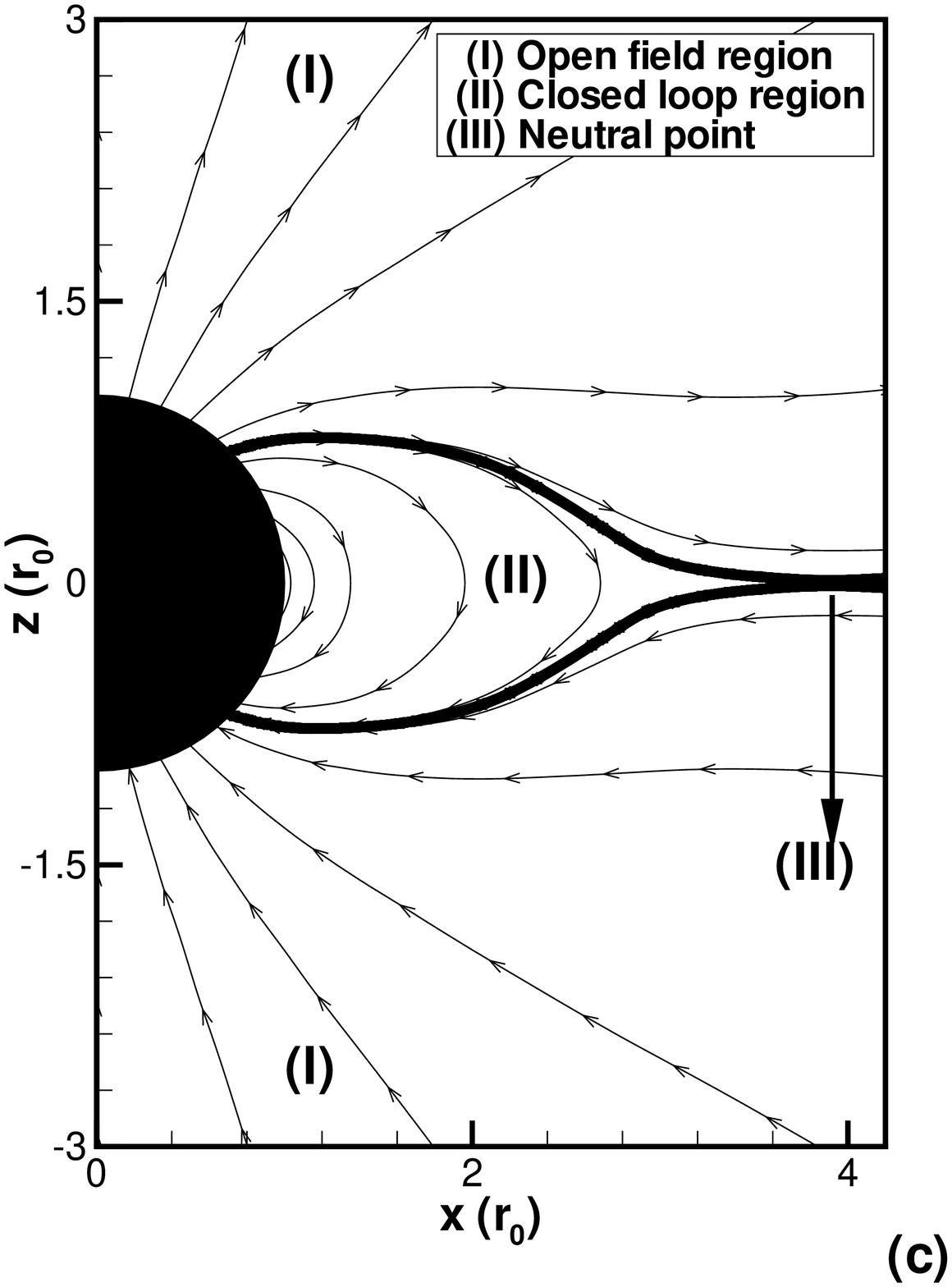}%
  \caption{(a) and (b): Evolution of the magnetic field lines for S03. The color of each streamline represents a given iteration, ranging from $N=1$ ($t=0$, blue) to $20,000$ ($t=7$h, black). At $N=1$, the field is a dipole. The steady-state is achieved at $N=148,000$ ($t=51$h). (c): Final steady-state configuration of the magnetic lines: (I) represents the zone of open field lines; (II) the zone of closed field lines; (III) the location of the neutral point. The thick line represents the position of the current sheet formed. \label{streamtraces2}}
\end{figure*}

Figures \ref{streamtraces2}a and \ref{streamtraces2}b present the temporal evolution of the magnetic field lines from iteration $N=1$ to $N=20,000$. This evolution was common to all runs. The plot presented is in the meridional plane, for S03. Figure~\ref{streamtraces2}a shows that at high latitudes, field lines tend to become open. Figure~\ref{streamtraces2}b shows the initial stretching of the lines emerging from low latitudes along the equator. 

\begin{figure}
\epsscale{1}
\plotone{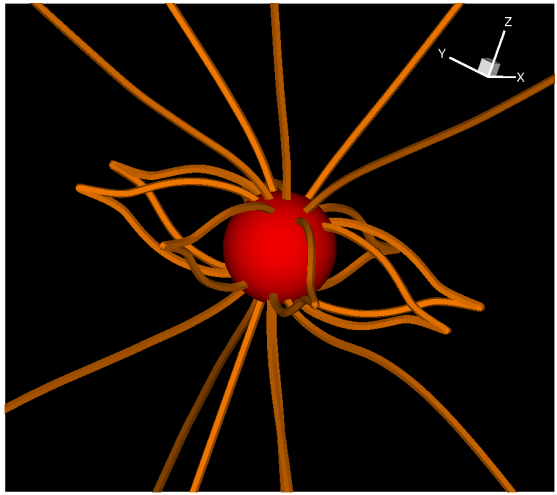}%
  \caption{3D view of the steady-state for S03. \label{3Dplot}}
\end{figure}

As the simulation evolves in time, the magnetic field lines reconnect in the equatorial plane. The steady-state configuration (Fig.~\ref{streamtraces2}c) consists of a formation of a bi-component wind. This configuration is composed by a zone of open magnetic field lines (I) coexisting with a zone of closed loops anchored on the star (II), also called {\it helmet streamer}. At the top of the closed loops lies the neutral point of cusp-type (III): approaching the neutral point from inside the closed loops, the magnetic field goes to zero. Beyond the neutral point, along the equatorial plane, a current sheet is formed (thick line in Fig.~\ref{streamtraces2}c). The zone of closed field lines is located at low latitudes from $\sim -45^o$ to $45^o$ and extends up to the neutral point. Beyond the neutral point, the zone of open field lines fills all the volume. Figure~\ref{3Dplot} shows a 3D view of the final configuration.

\subsubsection{Steady-State Wind Profile}
\begin{figure*}
\includegraphics[scale=0.26]{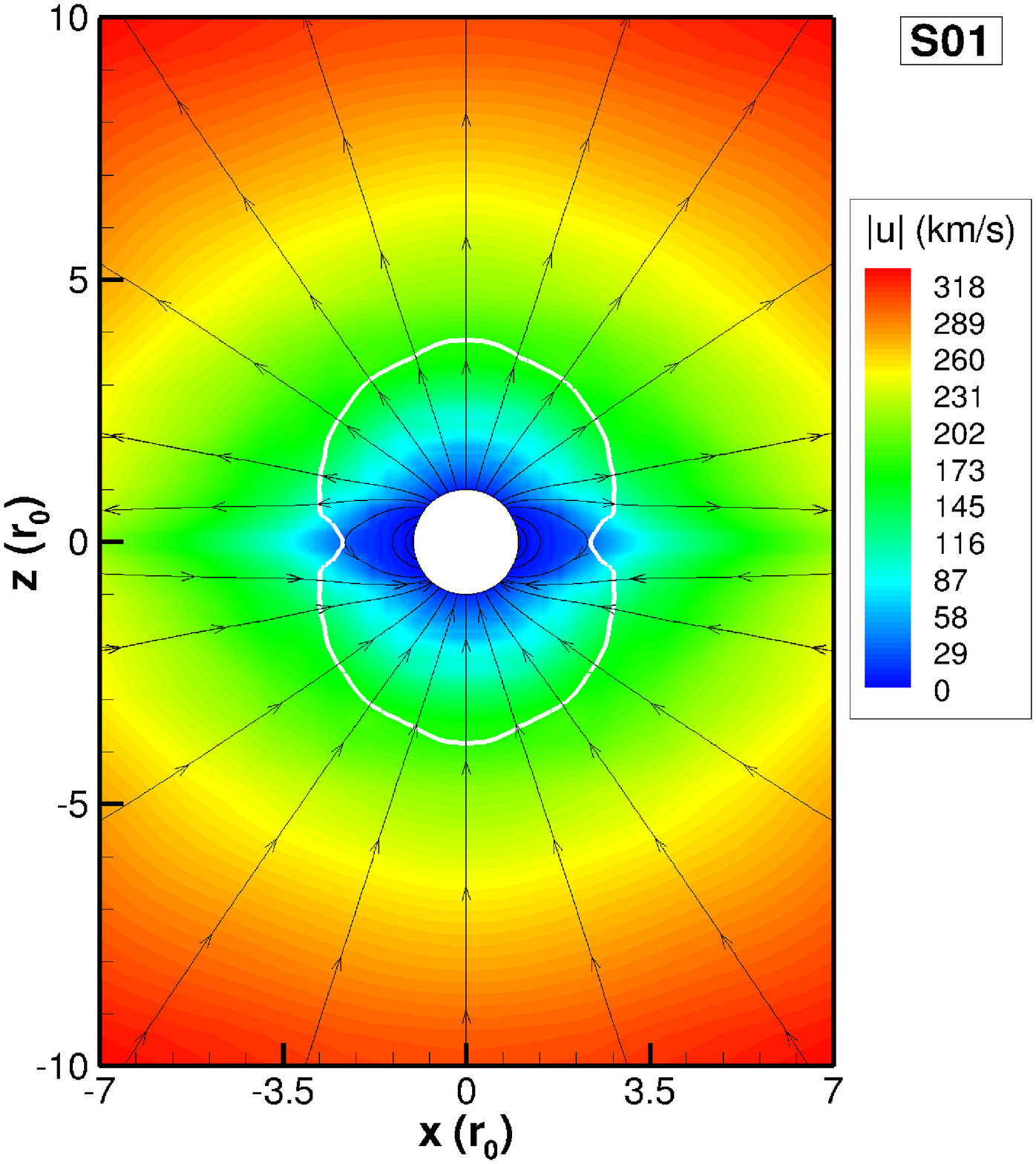}%
\includegraphics[scale=0.26]{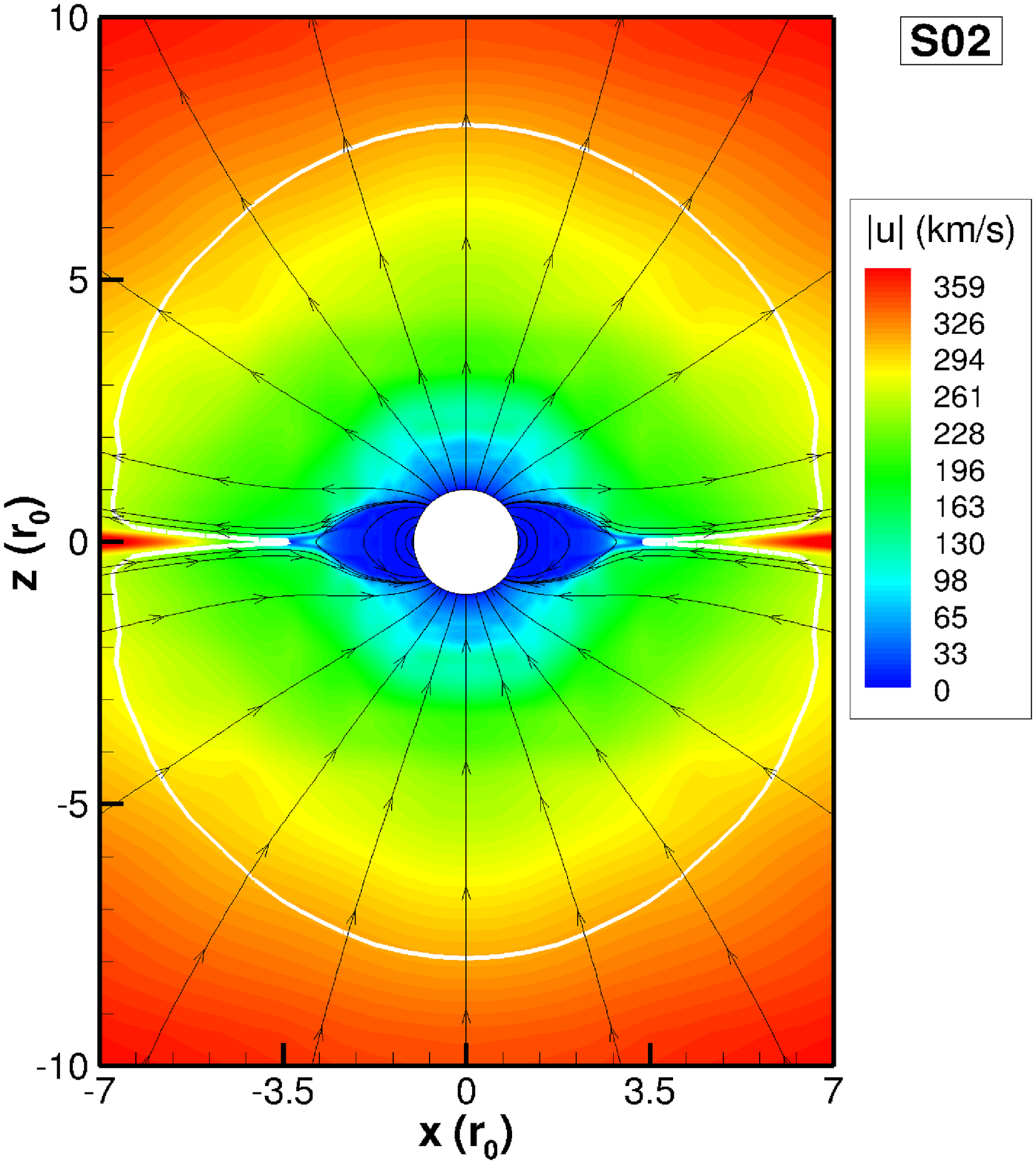}%
\includegraphics[scale=0.26]{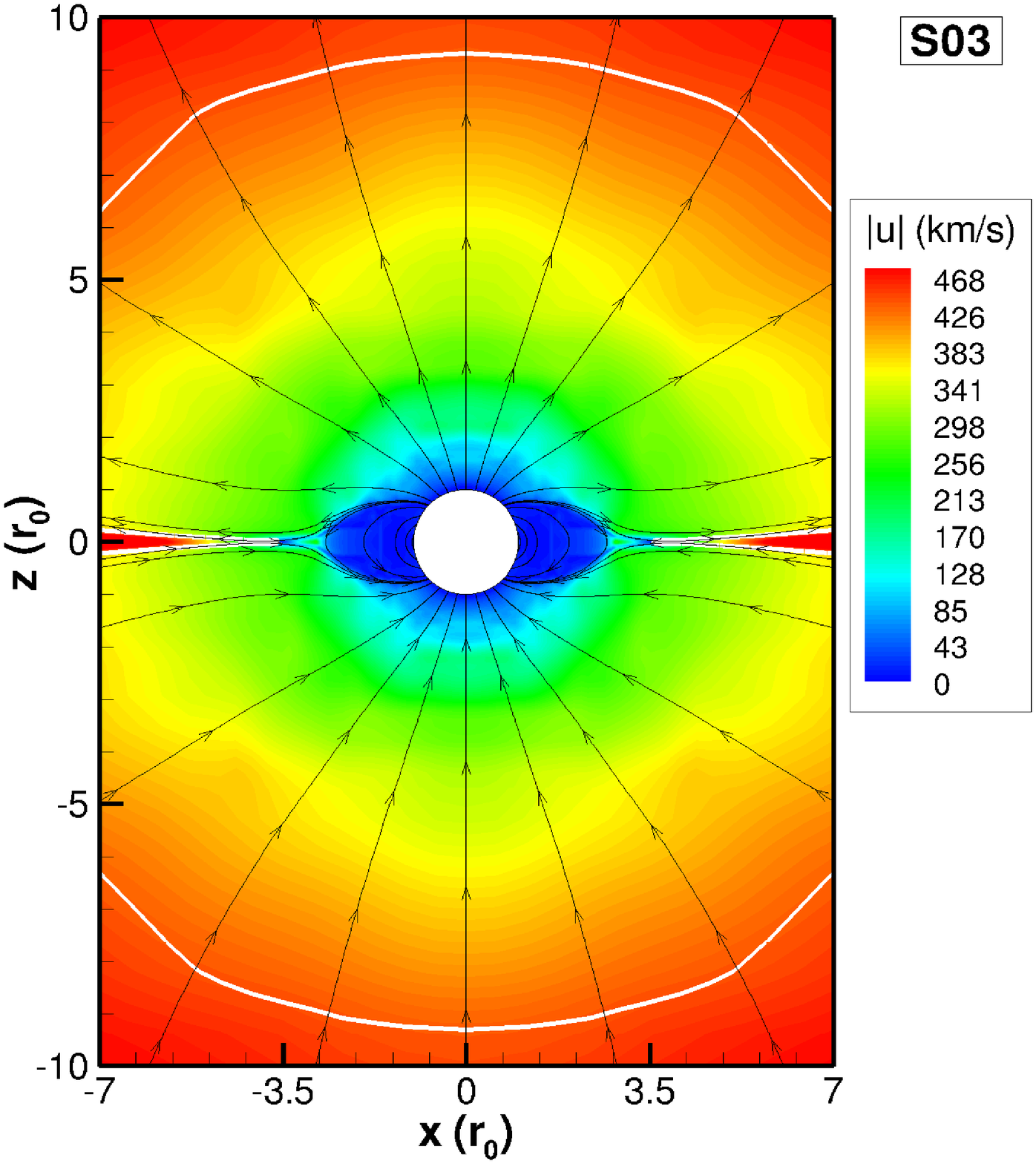}\\%
\includegraphics[scale=0.26]{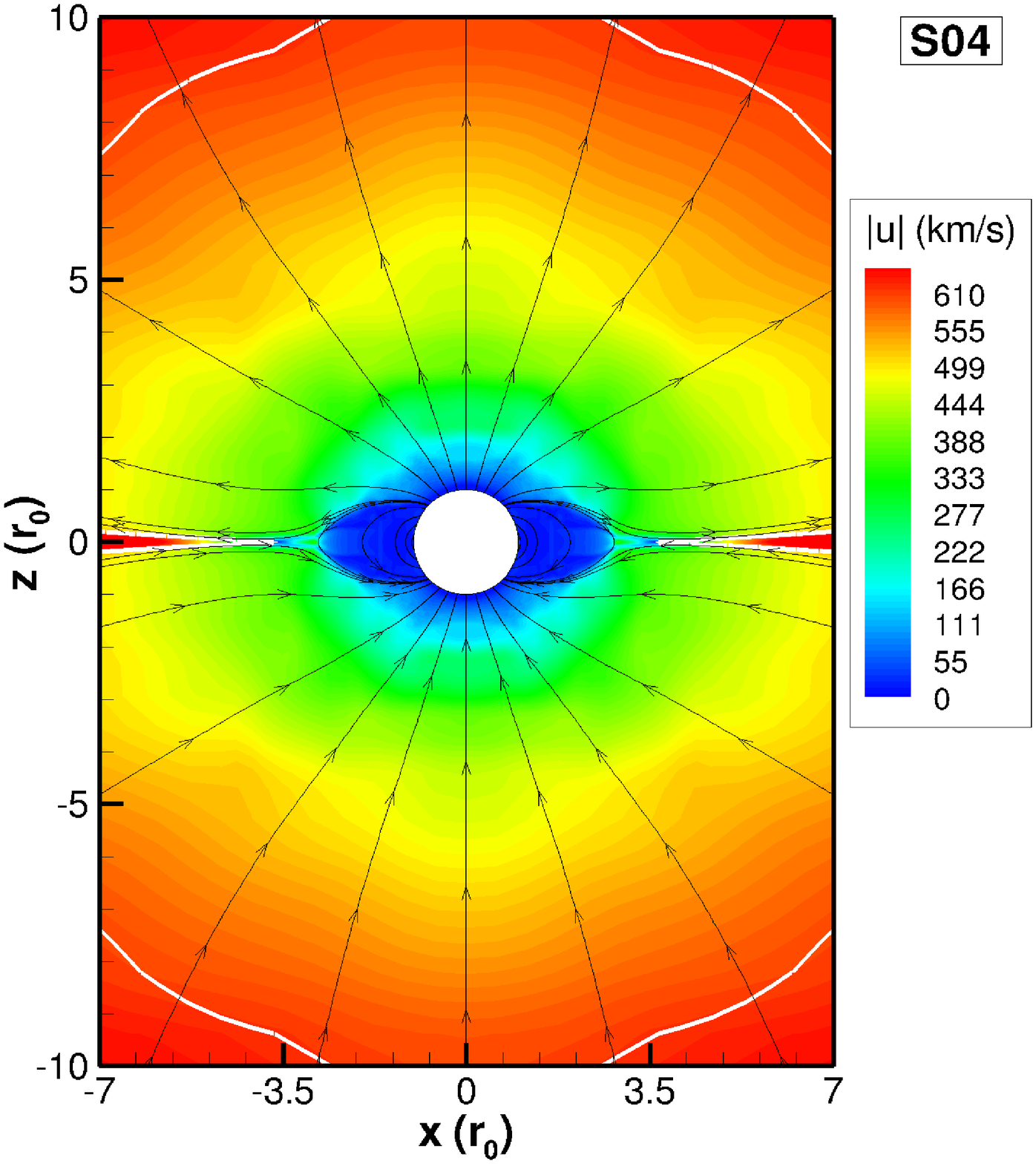}%
\includegraphics[scale=0.26]{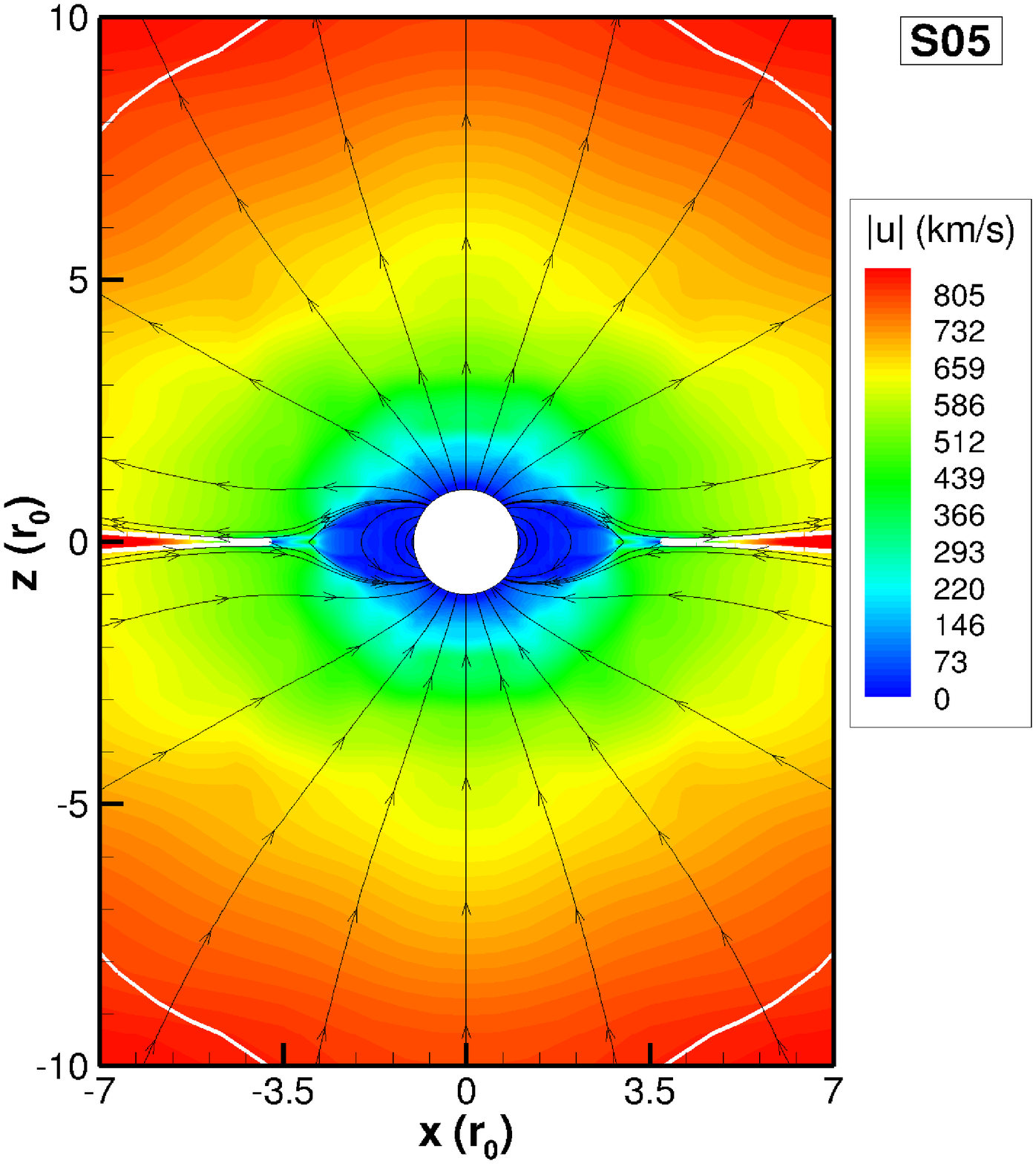}%
  \caption{Meridional cuts of the steady-state configurations for simulations S01 to S05. Black streamlines represent the final configuration of the magnetic field. Contour plots of the velocity of the flow are shown in the background. The white line is the Alfv\'en surface. \label{ss.firstset.utot}}
\end{figure*}

Figure~\ref{ss.firstset.utot} shows the total velocity of the wind in the meridional plane for the first set of simulations. The magnetic field lines are the black lines and the Alfv\'en surface is indicated by the white line. It can be seen that the increase of $B_0$ leads to faster winds.

The five panels present similar structures but quantitatively very different. The wind is not spherical; higher velocities are achieved at high latitudes. The higher $B_0$ is, the higher is the departure from spherical symmetry and the higher is the total velocity of the wind. 

The bi-modality of the wind is due to the nature of the magnetic force.  A purely HD (non-rotating) wind is spherically symmetric but in the MHD case, this symmetry is lost because the magnetic force has a meridional component. 

The solar wind has an observed bi-modal velocity ranging from $400$ to $800$~km~s$^{-1}$ at $1$~AU (solar minima). Although our model does not make use of magnetograms and thermal latitudinal heating as more realistic solar models such as \citet{2007ApJ...654L.163C} and \citet{2008ApJ...680..757L} do, it is interesting to compare S01 to their results. In \citet{2008ApJ...680..757L}'s model, at $r=4.5~r_0$, the highest velocity of $\sim 440$~km~s$^{-1}$ is achieved at the polar axis ($\theta=0^o$), while in the mid-latitudes  ($\theta \sim 45^o$), this value is decreased to $\sim 230$~km~s$^{-1}$. For S01, the same positions leads to $201$~km~s$^{-1}$ and $190$~km~s$^{-1}$, respectively. In our models, we are not treating the presence of waves and damping that is responsible for the latitude dependence that is observed in the solar wind. However, it is interesting that, just by the presence of magnetic field with an initially spherical wind, we obtain a latitude-dependent wind, although not as dramatic as in the Sun.

\begin{figure}
\epsscale{1}
\plotone{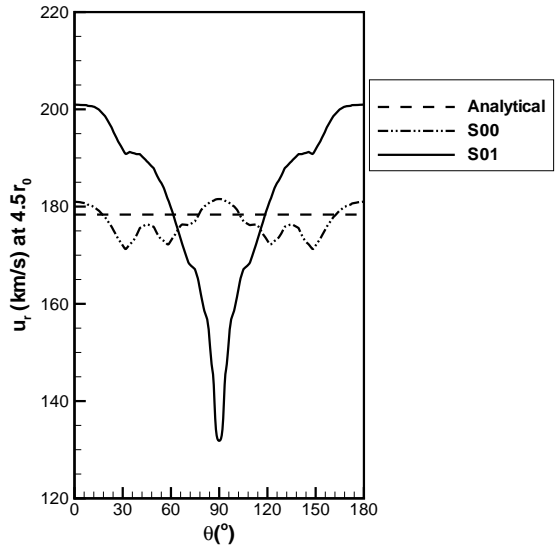}%
  \caption{Radial velocity of the wind as function of $\theta$ at $r=4.5~r_0$, for S00 (double-dot--dashed line), S01 (solid line) and analytical solution of a non-magnetized wind (dashed line).\label{uradial-s01}}
\end{figure}

Figure~\ref{uradial-s01} presents a radial cut at $r=4.5~r_0$ for cases S00 and S01. There is a range of angles, $\sim 30^o$ above and below the equator, where the MHD wind is slower compared to the HD model (S00). This deceleration is a consequence of the magnetic tension that is stronger near the equator. At high latitudes, the gradient of both thermal and magnetic pressures are responsible for driving the wind. Figure~\ref{uradial-s01} also presents the analytical result for a non-magnetized wind (dashed line) for comparison purposes. The width of the decrease seen in Fig.~\ref{uradial-s01} for the S01 curve is dependent on the grid resolution \citep{2004ApJ...611..575O}.

\begin{figure}
\epsscale{1}
\plotone{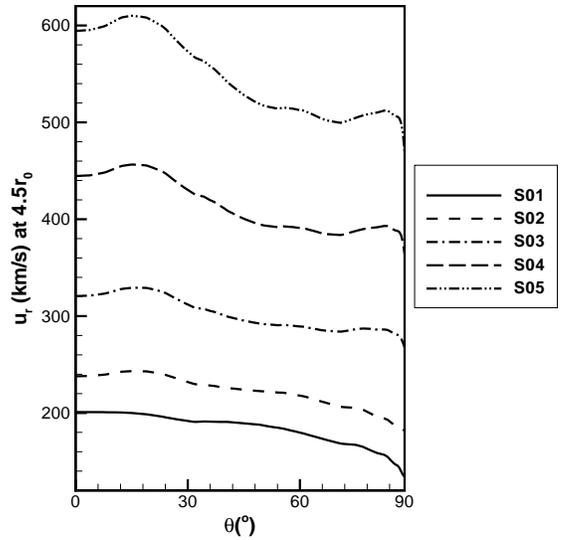}%
  \caption{Same as Fig.~\ref{uradial-s01}, but for S01 to S05, from $\theta=0^o$ to $\theta =90^o$.\label{variacaoUtot}}
\end{figure}

Figure~\ref{variacaoUtot} presents the radial cuts for the remaining simulations of the first set from $\theta=0^o$ to $90^o$. All the curves present smaller velocities at the equator ($\theta =90^o$) and higher velocities at high latitudes. The difference in velocities increases as $B_0$ is increased (see Table~\ref{table:4}). 

\begin{table}
\begin{center}
\caption{Co-latitude dependence of $u_r$ at $r=4.5~r_0$   \label{table:4}   }   
\begin{tabular}{c c c c}     
\tableline\tableline          
Name & $u_r (\theta = 18^o)$ & $u_r (\theta = 45^o)$ & $\Delta u_r$\tablenotemark{*} \\ 
\tableline    
S01 & $200 $  & $189$  & $11$  \\
S02 & $243$  & $224$   & $19$  \\
S03 & $329$  & $295$   & $34$  \\
S04 & $456$  & $400$   & $56$  \\
S05 & $610$   & $528$  & $82$  \\ 
\tableline                              
\end{tabular}
\tablenotetext{*}{$\Delta u_r = u_r (\theta = 18^o) - u_r (\theta = 45^o)$}
\end{center}
\end{table}

Another consequence of the latitudinal dependence of the Lorentz force, the mass-loss rate per unit solid angle of a steady wind 
\begin{equation}
\label{eq:mass-loss}
\dot{m}(\theta) = r^2 \rho(\theta) u_r(\theta)
\end{equation}
is also latitude-dependent. According to Eq.~(\ref{eq:continuity_conserve}), $\dot m(\theta)$ is constant if there is no variation in $\theta$. In our simulations, this is verified far from the central star. Table~\ref{table:3} presents the calculated $\dot m(\theta)$ for two given co-latitudes of the steady-state wind, when it has already reached constant values. $\dot m(\theta)$ at the pole is smaller compared to other angles, despite the increase of the radial velocity. The lower value of $\dot m(\theta)$ at the pole is due to matter carried from high latitudes to low latitudes (\S\ref{subsec.meridional}). As the latitudinal increase of density is higher than the corresponding decrease in the radial velocity, $\dot m(\theta)$ is higher at low latitudes. 

\begin{table}
\begin{center}
\caption{Mass-loss rate per unit solid angle of the outflow\tablenotemark{*} \label{table:3}}   
\begin{tabular}{c c c}     
\tableline\tableline          
Name & $\dot{m}_{15} (\theta = 0^o)$ & $\dot{m}_{15} (\theta = 45^o)$ \\ 
\tableline    
S02 & $5.1$  & $5.7$ \\
S03 & $6.0$ & $6.6$ \\
S04 & $7.1$  & $7.5$ \\
S05 & $8.2$   & $8.7$ \\ 
\tableline                              
\end{tabular}
 \tablenotetext{*}{$\dot{m}_{15}(\theta) = [\dot{m}(\theta) / 10^{-15}]~M_\odot$~yr$^{-1}$}
\end{center}
\end{table}

\subsubsection{Meridional Flows}\label{subsec.meridional}
The component of the magnetic force in the $\theta$ direction gives rise to meridional flows, bringing matter from both hemispheres towards the equator. This causes a density enhancement along the equatorial plane.

\begin{figure*}
\includegraphics[scale=0.26]{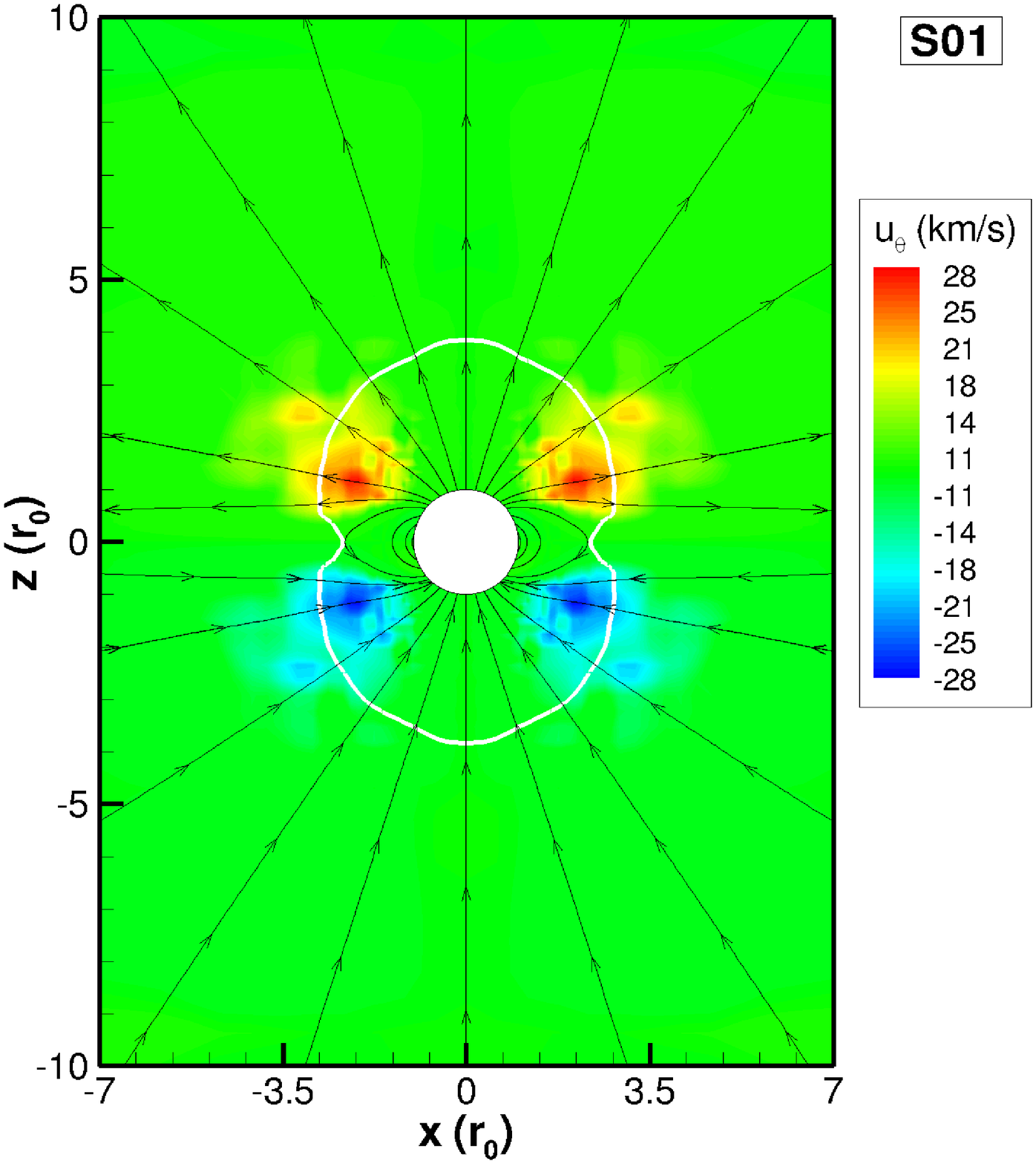}%
\includegraphics[scale=0.26]{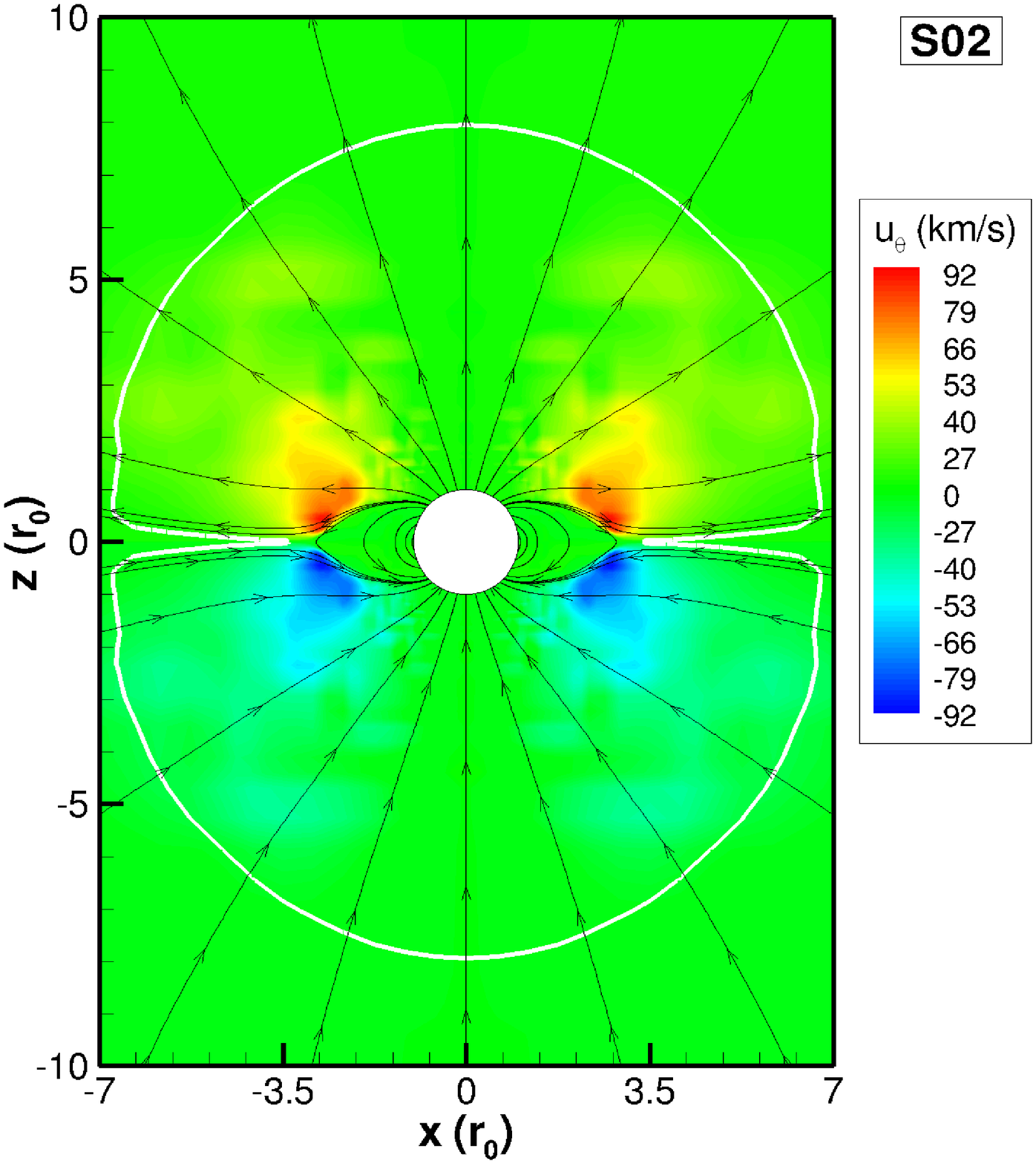}%
\includegraphics[scale=0.26]{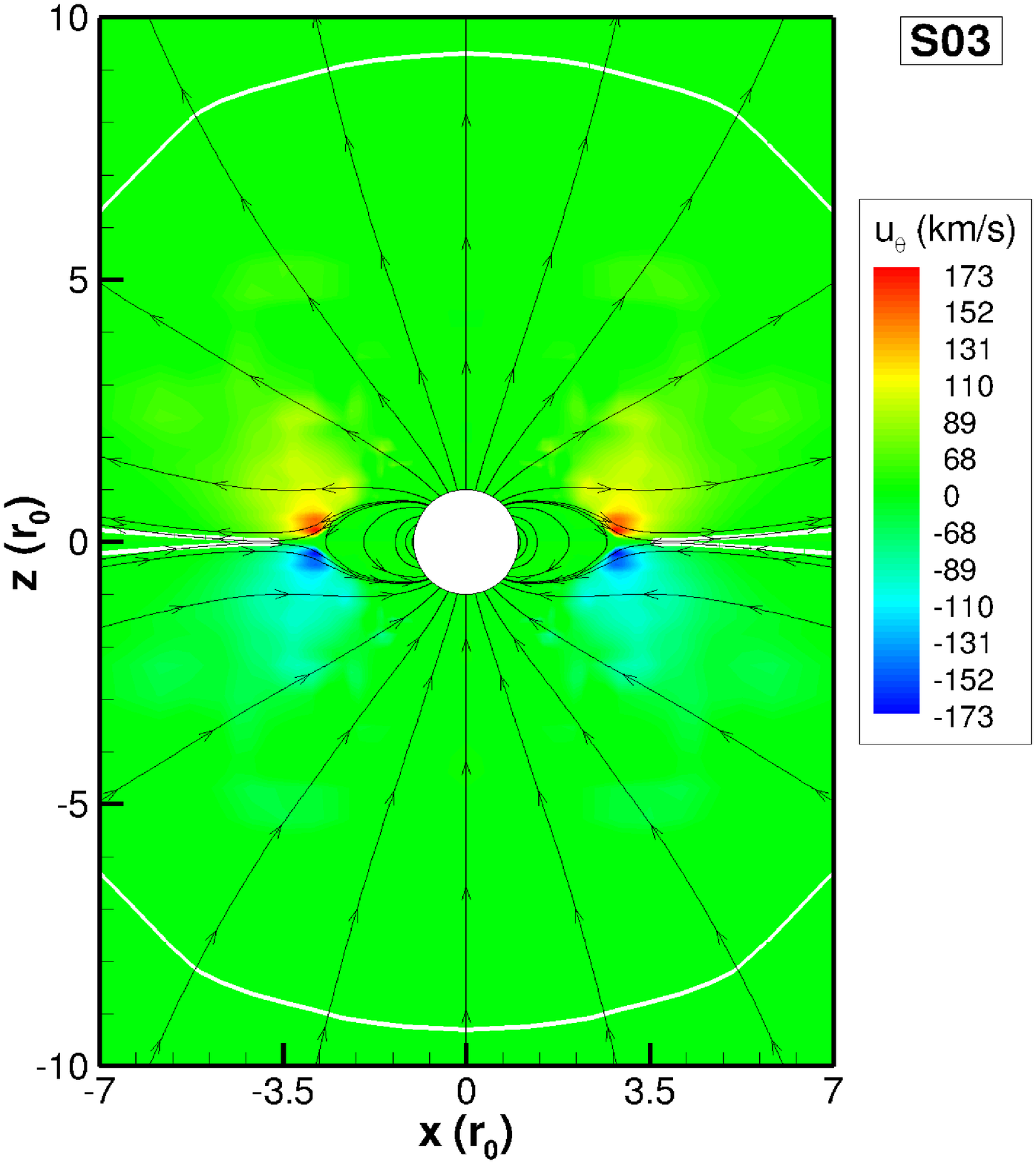}\\%
\includegraphics[scale=0.26]{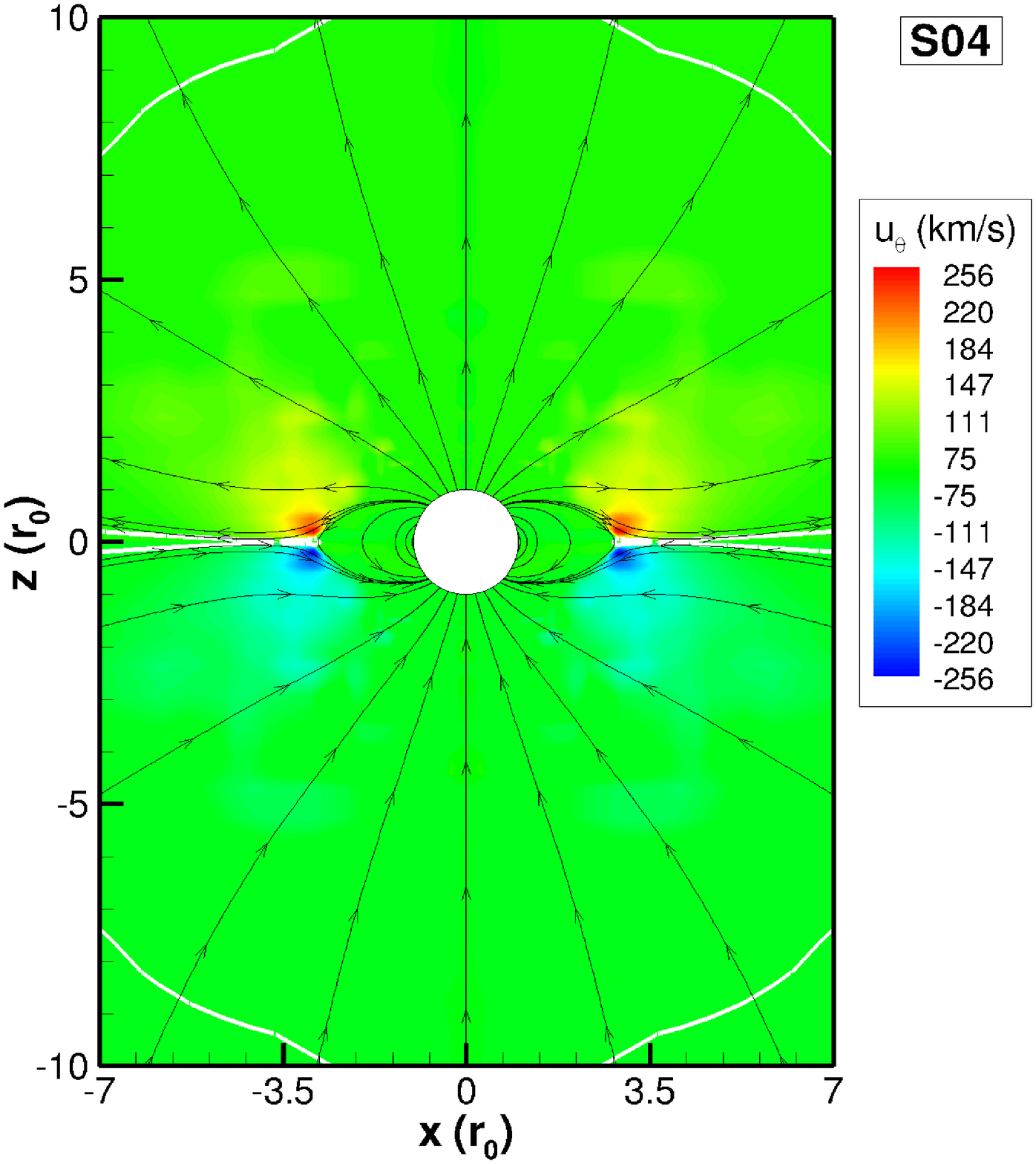}%
\includegraphics[scale=0.26]{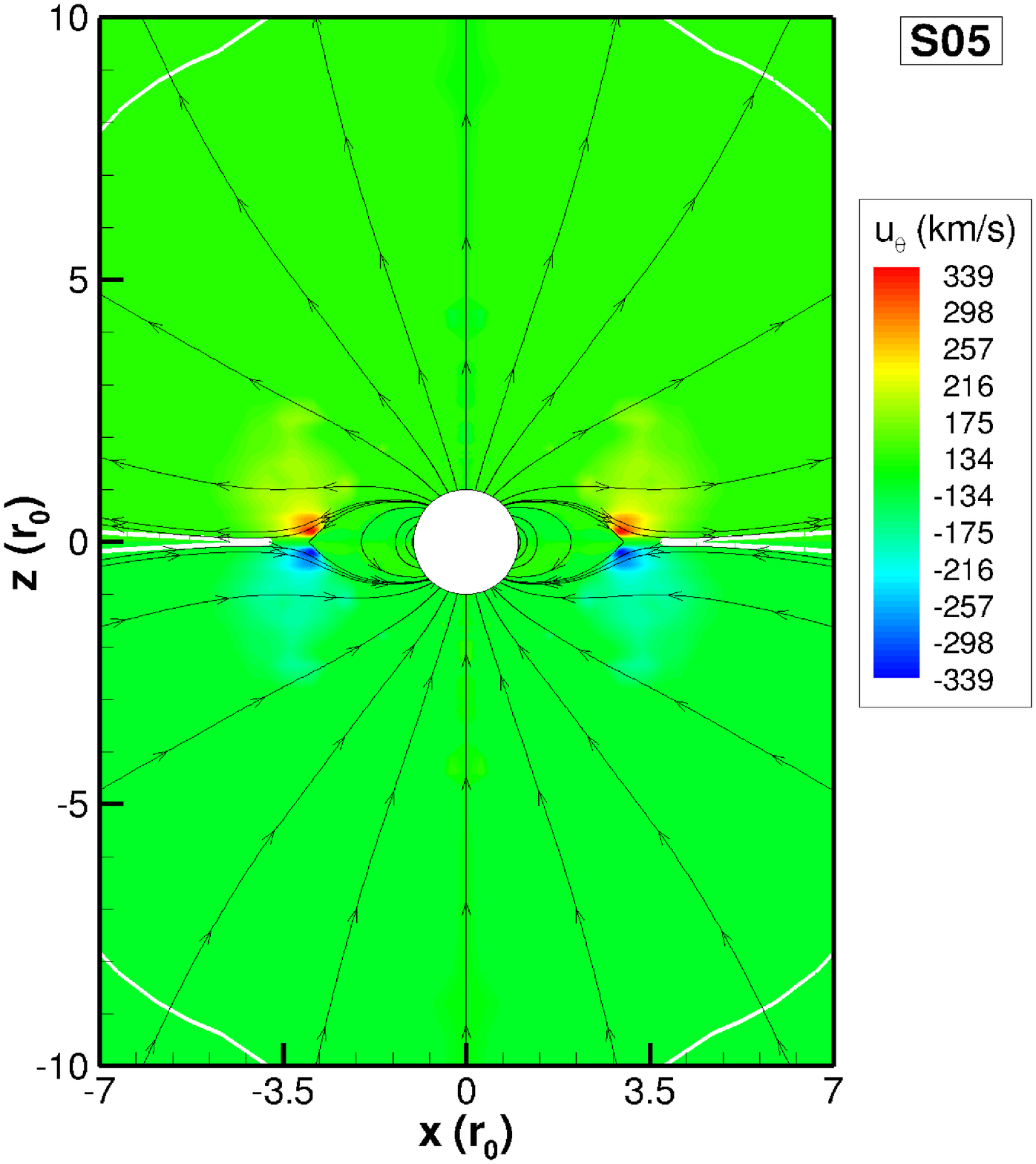}%
\includegraphics[scale=0.26]{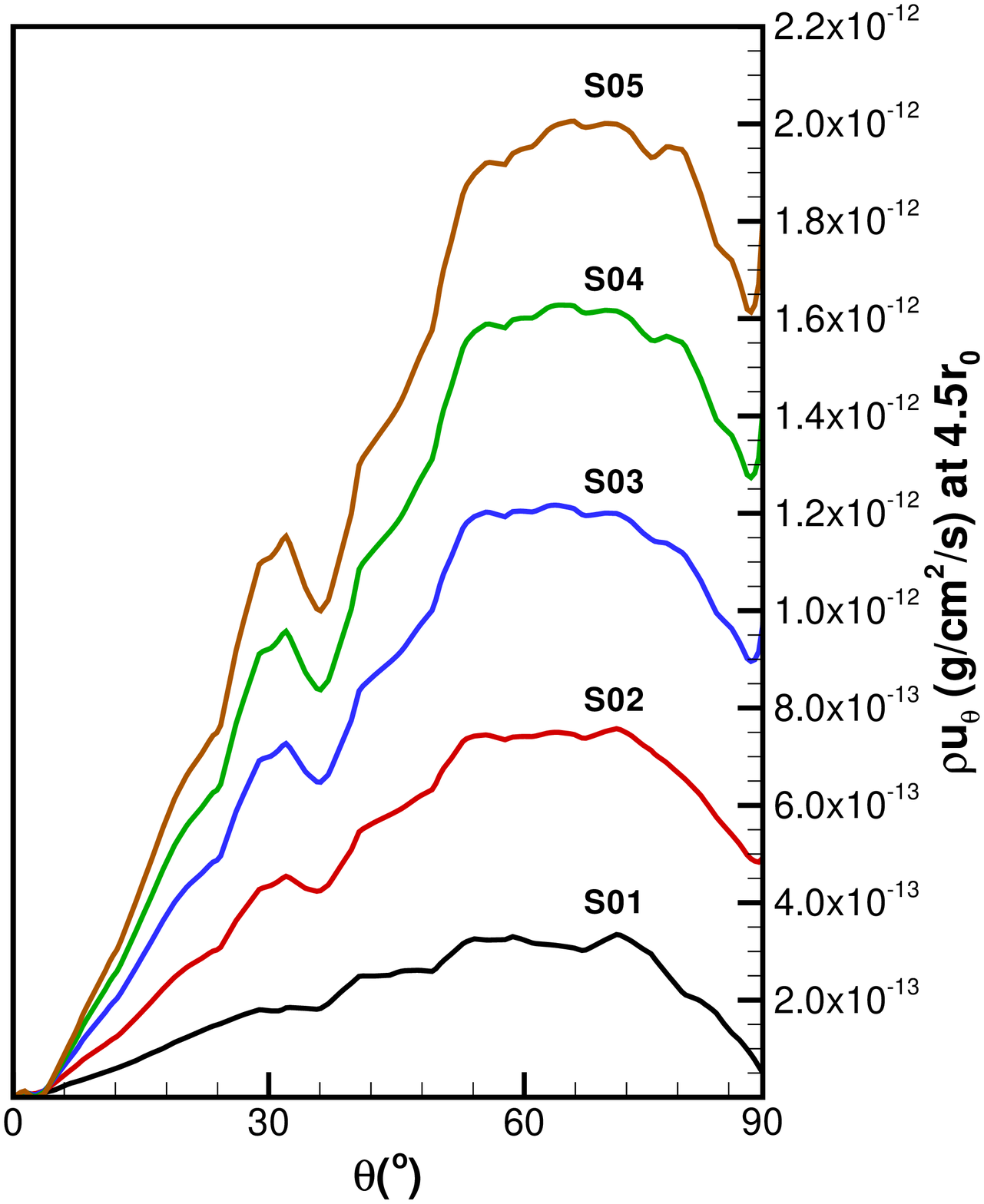}%
  \caption{Meridional velocities $u_\theta$ for the steady-state configurations of simulations S01 to S05. Black streamlines are the magnetic field lines. The white line is the Alfv\'en surface. Last panel: The flux of matter in the $\theta$ direction, $\rho u_\theta$, for S01 to S05 at $r=4.5~r_0$.  \label{ss.firstset}}
\end{figure*}

Figure~\ref{ss.firstset} presents the meridional velocity, $u_\theta$, for S01 to S05 at steady-state. $u_\theta$ increases with $B_0$, ranging from a maximum value of $28$~km~s$^{-1}$ in S01 ($B_0=1$~G) to $339$~km~s$^{-1}$ in S05 ($B_0=20$~G). The last panel of Fig.~\ref{ss.firstset} presents the flux of matter in the $\theta$ direction, $\rho u_\theta$, at $4.5~r_0$. It can be seen that as the magnetic field increases, there is an increase of the meridional flux of matter. 

\subsubsection{The Neutral Point}
The meridional flows are also responsible for compressing lines with opposite polarities along the equatorial plane and due to reconnection, a current sheet is formed. In our simulations, the reconnection is controlled due to numerical resistivity. In order to minimize the effects of numerical resistivity, we refined the grid along the equatorial plane. Ultimately, a more realistic reconnection should be included in the global 3D MHD simulations such as done by \citet{2007JGRA..11210210K}. The reconnection leads to the emergence of a faster wind in the equatorial plane, beyond the reconnection point (Fig.~\ref{ss.firstset.utot}). Such wind is an artifact of the model and the same behavior was also observed by \citet{2003ApJ...595L..57R}. 

The location of the neutral point $r_N$ is given by the requirement that the total pressure (magnetic plus thermal) at the tip of the cusp of closed loops is continuous 
\begin{equation}\label{eq.neutralpoint}
p(r_N^+) - p(r_N^-) = \frac{B(r_N)^2}{8\pi} \, , 
\end{equation}
where $p(r_N^+)$ is the thermal pressure evaluated immediately after $r_N$ and $p(r_N^-)$ immediately before $r_N$ \citep{1971SoPh...18..258P}. Along a magnetic streamline, the magnetic force is null. Assuming an isothermal atmosphere, the energy equation reduces to 
\begin{equation}\label{eq.iso.bernoulli}
\frac{u^2}{2} + c_s^2 \ln \frac{p}{p_0} + g r = \frac{u_0^2}{2} + g_0 r_0 \, ,
\end{equation}
where $c_s$ denotes the isothermal sound speed and the use of the subscript ``0'' indicates that the quantity is evaluated at the base of the wind. Equation~(\ref{eq.neutralpoint}) can be written as
\begin{eqnarray}\label{eq.neutralpoint2}
p(r_N^+) - p(r_N^-) \simeq p(r_N^+) \left[ \exp \frac 12 \left( \frac{u(r_N)}{c_s} \right)^2  -1 \right] \nonumber \\ \simeq p(r_N^+) \frac{u(r_N)^2}{2 c_s^2}\, , 
\end{eqnarray}
where we neglected the initial velocity of the flow in comparison with the velocity of the flow at the neutral point, $u(r_N)$, and assumed $u(r_N) \ll c_s$. Under these considerations, Eqs.~(\ref{eq.neutralpoint}) and (\ref{eq.neutralpoint2}) lead to 
\begin{equation}
\frac{B(r_N)^2}{8\pi} \simeq p(r_N^+) \frac{u(r_N)^2}{2 c_s^2} \simeq \frac 12 \rho(r_N^+) {u(r_N)^2} \, ,
\end{equation}
implying that at the neutral point
\begin{equation}
{u(r_N)} \simeq {v_A(r_N)}\, ,
\end{equation}
where $v_A$ is the Alfv\'en velocity. I. e., the neutral point lies on the Alfv\'en surface. 

Figure~\ref{ss.firstset} presents the Alfv\'en surface (white line), where it can be seen that the above result is confirmed by our simulations. In Table~\ref{table:2} the position of the neutral point $r_N$ is given for each of the simulations of the first set. The cell size at the current sheet is $ 0.036~r_0$, so the numerical error associated with $r_N$ is $\pm 0.018~r_0$. $r_N$ increases as $B_0$ is increased, i. e., the ``dead-zone'' (zone of closed loops) is larger for larger $B_0$. As $B_0$ is increased from $5$~G to $20$~G, $r_N$ moves from $3.68~r_0$ to $4.12~r_0$. This is expected, since the Alfv\'en surface moves farther from the star as $B_0$ is increased. S01 does not present reconnection. 

\begin{table}
\begin{center}
\caption{Location of the neutral point, $r_N$, along the equatorial plane as a function of $B_0$ \label{table:2}   }   
\begin{tabular}{c c c}     
\tableline\tableline          
Name & $B_0$ (G) & $r_N (r_0)$ \\ 
\tableline    
S02 & $5 $  & $3.68$ \\
S03 & $10$  & $3.90$ \\
S04 & $15$  & $4.01$ \\
S05 & $20$  & $4.12$ \\ 
\tableline                              
\end{tabular}
\end{center}
\end{table}

\begin{figure}
\epsscale{1}
\plotone{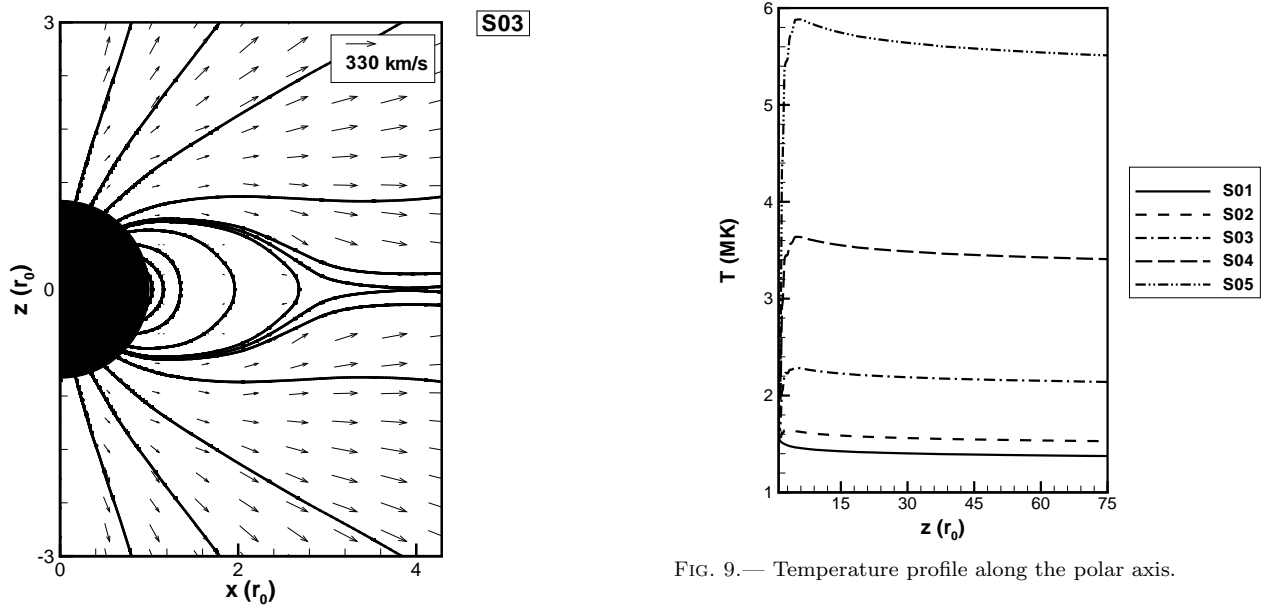}
  \caption{Trapping of particles inside the closed loop region for S03. Thick lines represent the magnetic field lines and vectors represent the flow velocity, both in the meridional plane.\label{trapping}}
\end{figure}

Inside the closed field line region, particles are trapped. This is evidenced in Fig.~\ref{trapping}, where we plot the magnetic field lines and the vectors of the flow velocity. As can be seen, if a given particle emerges inside the closed loop zone, it will remain there, because its velocity is not high enough to escape the effective potential well created by the magnetic tension and gravitational attraction force. 

\subsubsection{Energetics}
We would like to estimate the contribution of the different energies to the total energy content of the system.

\begin{figure}
\epsscale{1}
\plotone{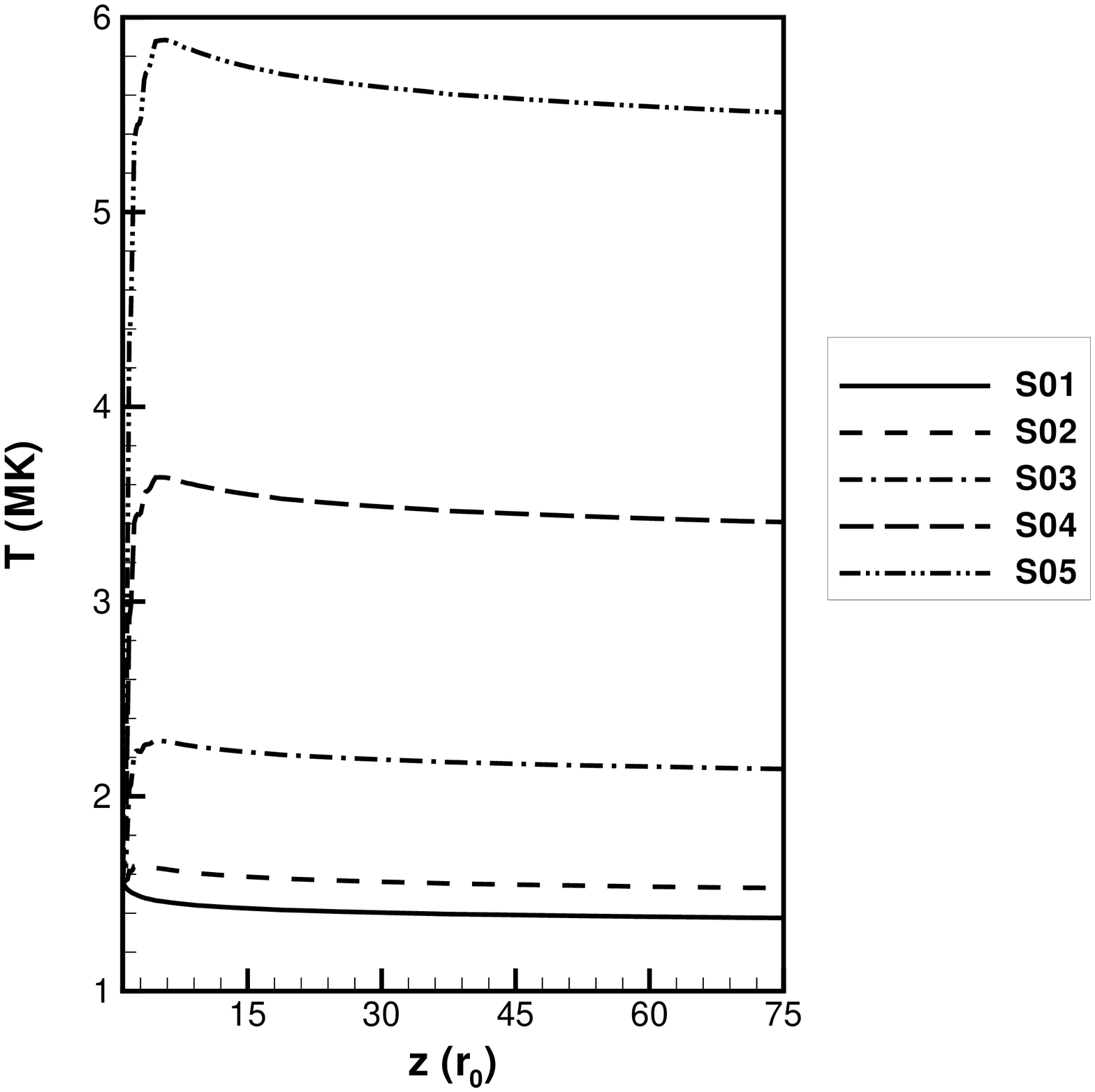}%
  \caption{Temperature profile along the polar axis.\label{fig.temperature}}
\end{figure}

The temperature $T$ of the steady-state coronal wind increases with $B_0$. Figure~\ref{fig.temperature} presents the temperature profiles along the polar axis for the cases of the first set of simulations (in steady-state). For higher $B_0$, $T$ increases near the base of the wind and it drops slowly for larger radii. For S05, the increase in the temperature is maximum among all the cases, reaching $\simeq 5.9$~MK around $6~r_0$. In the solar wind, however, the temperature profile is not flat \citep{2006JGRA..11110103M}, because the bulk of the internal energy deposition occurs near the Sun.

For steady-state, conservation of total energy flux is given by
\begin{equation}\label{eq:div.q}
{\bf \nabla} \cdot {\bf q} = 0 \, ,
\end{equation}
where ${\bf q}$ is the total energy flux given by
\begin{equation}\label{eq.q}
{\bf q} = {\bf u} \left( \frac{\rho u^2}{2} + \frac{\gamma p}{\gamma -1} + \rho \Phi \right) + {\bf S} \, ,
\end{equation}
and $\Phi$ is the gravitational potential. The terms on the right hand side of Eq.~(\ref{eq.q}) are the kinetic, enthalpy, and gravitational energy fluxes, and ${\bf S}$ is the Poynting flux vector that in ideal MHD is given by
\begin{equation}\label{eq:poynting.MHD}
{\bf S} = \frac{B ^2}{4\pi} {\bf u} -  \frac{({\bf u }\cdot {\bf B})}{4\pi} {\bf B} \, .
\end{equation}

\begin{figure}
\epsscale{.5}
\plotone{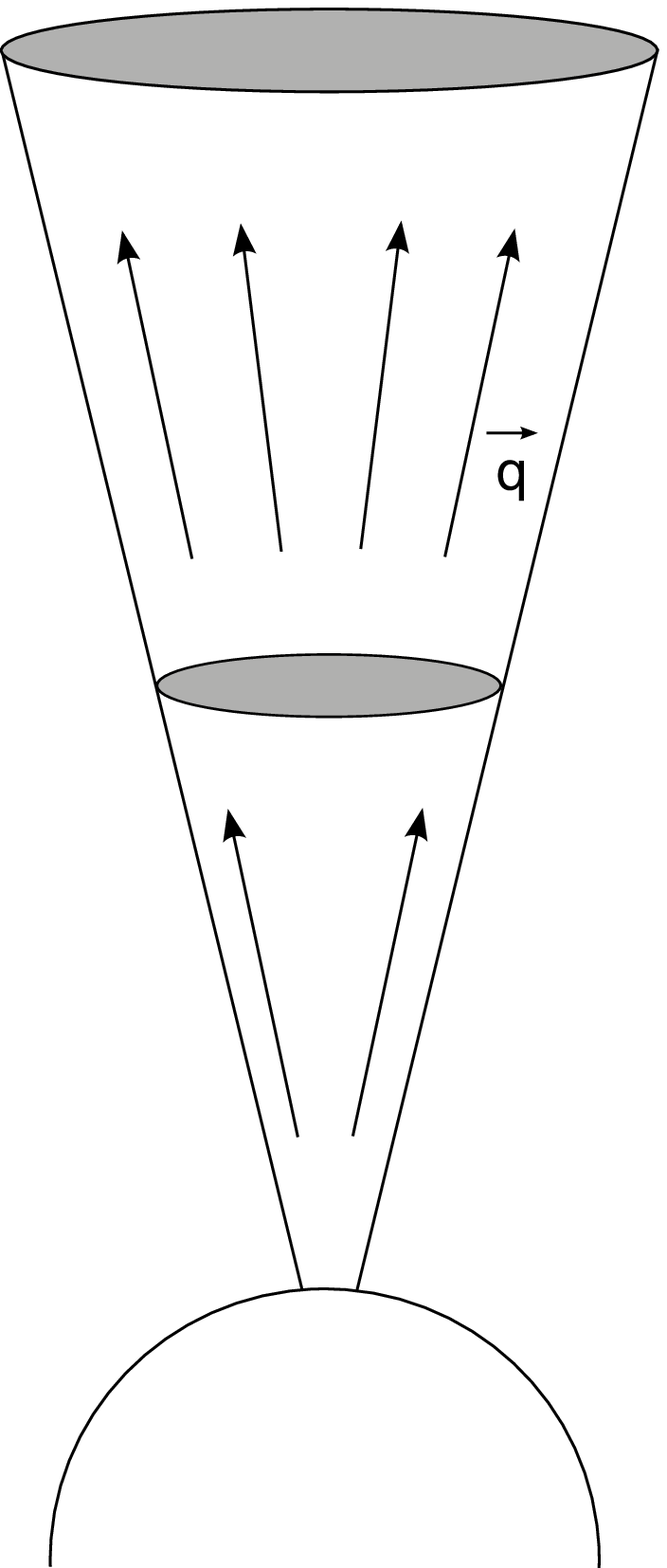}%
  \caption{Cartoon of the energy flux  ${\bf q}$ along a flux tube. \label{sketch-flux-tube}}
\end{figure}

Taking a volume defined by a given magnetic flux tube bounded by two cross-sections (see Fig.~\ref{sketch-flux-tube}), the net energy flux ${\bf q}$ should be conserved.

Table~\ref{table:fluxtube} presents each parcel of the total energy flux that crosses four cross-sectional areas of a magnetic flux tube whose central axis is located at $\theta \sim 30^o$. At $2.2~r_0$, the enthalpy power (i. e., the thermal plus internal energy fluxes integrated over the area of the cross-section of the tube at this position) is already the dominant form of energy. At $3.0~r_0$, magnetic power is almost negligible indicating that the continuous acceleration of the flow from there on is driven by solely the enthalpy. At $r=4.2~r_0$ and $6.0~r_0$, the conversion of enthalpy into kinetic energy is responsible for the continuous acceleration of the wind until a terminal velocity is reached.  

\begin{deluxetable}{l r r r r }  
\tablewidth{0pt}
\tablecaption{Percentage amount of energy flux passing through four different cross-sections along a flux tube for S03 and S05 \label{table:fluxtube}  }   
\tablehead{\colhead{Energy Flux} & \colhead{$2.2~r_0$} &\colhead{$3~r_0$} &\colhead{$4.2~r_0$} &\colhead{$6~r_0$} }
\startdata
\cutinhead{S03}
Magnetic	&	$	0.15	$	&	$	0.07	$	&	$	0.02	$	&	$	0.01	$	\\
Kinetic	&	$	0.32	$	&	$	0.59	$	&	$	1.26	$	&	$	1.76	$	\\
Enthalpy	&	$	97.08	$	&	$	97.63	$	&	$	97.50	$	&	$	97.39	$	\\
Grav.	&	$	2.45	$	&	$	1.72	$	&	$	1.22	$	&	$	0.85	$	\\
\cutinhead{S05}
Magnetic	&	$	0.41	$	&	$	0.15	$	&	$	0.02	$	&	$	0.02	$	\\
Kinetic	&	$	0.46	$	&	$	0.85	$	&	$	1.77	$	&	$	2.33	$	\\
Enthalpy	&	$	98.04	$	&	$	98.27	$	&	$	97.72	$	&	$	97.31	$	\\
Grav.	&	$	1.09	$	&	$	0.72	$	&	$	0.50	$	&	$	0.34	$	
\enddata
\end{deluxetable}

\subsection{The Second Set of Simulations}
The second set of simulations aims to investigate the effect of the density on the wind. We vary the initial density maintaining the same surface magnetic field $B_0$ and surface temperature $T_0$ adopted in S05. The simulations we analyze in this section consist of simulations S06 to S09 (Table~\ref{table:1}).

\begin{figure}
\epsscale{1}
\plotone{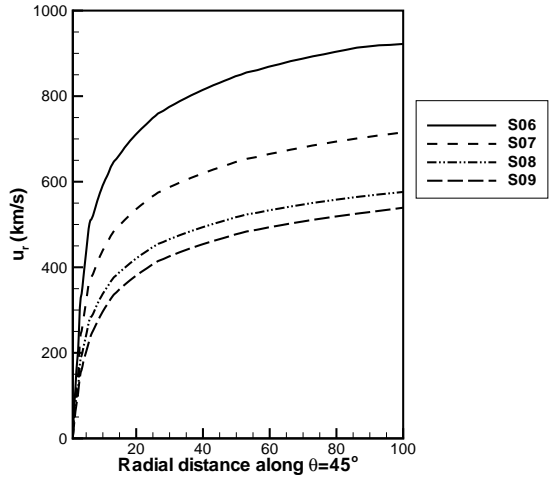}
  \caption{Radial velocity profiles as a function of the distance along $\theta =45^o$ for the second set of simulations.\label{ss.uradial-set02}}
\end{figure}

We note that as $\rho_0$ increases, there is a decrease in the wind velocity, as can be seen in Fig.~\ref{ss.uradial-set02}, where we plotted the radial velocity of the flows along $\theta =45^o$. Also, as $\rho_0$ is increased, the reconnection point moves closer to the star. 

\subsection{Comparison Between First and Second Sets}\label{sec.comp}
Comparing the first and the second set of simulations, we note that flows with same $\beta_0$ possess the same wind velocity profiles (S01/S09, S02/S08, S03/S07, and S04/S06). This result can be verified from Eqs.~(\ref{eq:continuity_conserve}) to (\ref{eq:energy_conserve}). The magnetic field can be written in terms of a normalized field $\tilde{B}$ with a normalization constant $B_0$
$$ {\bf B} = {\bf \tilde{B}} B_0 \, .$$
Doing the the same for the density
$$ \rho = \tilde \rho \rho_0 \, .$$
In this case, Eqs(\ref{eq:continuity_conserve}) to (\ref{eq:energy_conserve}) are written as
\begin{equation}
\label{eq:continuity_conserve3}
\frac{\partial \tilde\rho}{\partial t} + \nabla\cdot \left(\tilde \rho {\bf u}\right)  = 0
\end{equation}
\begin{equation}
\label{eq:momentum_conserve3}
\frac{\partial \left(\tilde \rho {\bf u}\right)}{\partial t} + \nabla\cdot\left[ \tilde\rho{\bf u\,u}+ \left(\tilde p + \frac{B_0^2}{\rho_0} \frac{\tilde{B}^2}{8\pi}\right)I - \frac{B_0^2}{\rho_0} \frac{{\bf \tilde B\,\tilde B}}{4\pi}\right]= \tilde\rho {\bf g}
\end{equation}
\begin{equation}
\label{eq:bfield_conserve3}
 \frac{\partial {\bf \tilde B}}{\partial t} + \nabla\cdot\left({\bf u\,\tilde B} - {\bf \tilde B\,u}\right)  = 0
\end{equation}
\begin{equation}
\label{eq:energy_conserve3}
\frac{\partial\tilde \varepsilon}{\partial t} +  \nabla \cdot \left[ {\bf u} \left( \tilde \varepsilon + \tilde p + \frac{B_0^2}{\rho_0}\frac{\tilde B^2}{8\pi} \right) - \frac{B_0^2}{\rho_0}\frac{\left({\bf u}\cdot{\bf \tilde B}\right) {\bf \tilde B}}{4\pi}\right]= \tilde \rho {\bf g}\cdot {\bf u} \, ,
\end{equation}
where
\begin{equation}
\tilde \varepsilon=\frac{\tilde \rho u^2}{2}+\frac{\tilde p}{\gamma-1}+\frac{B_0^2}{\rho_0}\frac{\tilde B^2}{8\pi} \, .
\end{equation} 
Equations~(\ref{eq:continuity_conserve3}) and (\ref{eq:bfield_conserve3}) show that the normalized density profile ($\tilde \rho$) and the normalized field profile ($\tilde B$) are independent of the values $\rho_0$ and $B_0$ adopted at the surface. Equations (\ref{eq:momentum_conserve3}) and (\ref{eq:energy_conserve3}) show that the choice of $\rho_0$ and $B_0$ (or in other words the choice of $\beta_0 \propto \rho_0/B_0^2$, as we assumed the same base temperature) appears explicitly in the momentum and energy equations. According to Eqs.~(\ref{eq:continuity_conserve3}) to (\ref{eq:energy_conserve3}), it is expected that flows with same $\beta_0$ have the same wind velocity, magnetic field configuration, density profile, etc. Although the normalized density and magnetic field profiles are the same if $\beta_0$ is the same in two different simulations, the mass-loss rates are different, where the value of the normalization constant $\rho_0$ has to be considered. 

Hence, by analyzing only the velocity of an outflow, one cannot equivocally predict its physical characteristics at its base. Our results indicate that there is a group of magnetized flows that would present the same terminal velocity despite of its thermal and magnetic energy densities, as long as $\beta_0$ is the same.

This degeneracy can be removed if we consider the mass-loss rate. As the velocity profiles for two different magnetized outflows with similar $\beta_0$ are the same, the difference between the mass-loss rates of these winds comes from different density structures. To illustrate, consider cases S01 and S09 with $\beta_0=1$: both cases present the same velocity profile and magnetic field configuration (although the intensity is not equal); but the density, and consequently the mass-loss rate, of S09 is 400 times larger than S01 in the entire numerical domain.

\subsection{The Third Set of Simulations With $\gamma = 1.1$}
Throughout the present paper, we have assumed that $\gamma = 1.01$. However, more realistic magnetized wind models compute the heating of the wind caused due to physical processes, as is the case of the 1D models done by \citet{2006ApJ...639..416V, 2006MNRAS.368.1145F, 2007ApJS..171..520C}, among others. However, inclusion of such processes, e.g. dissipation of waves or turbulence, in a 3D code is very challenging. We thus opted to parametrize the energy content of the wind by the use of $\gamma$.

\begin{figure}
\includegraphics[scale=0.26]{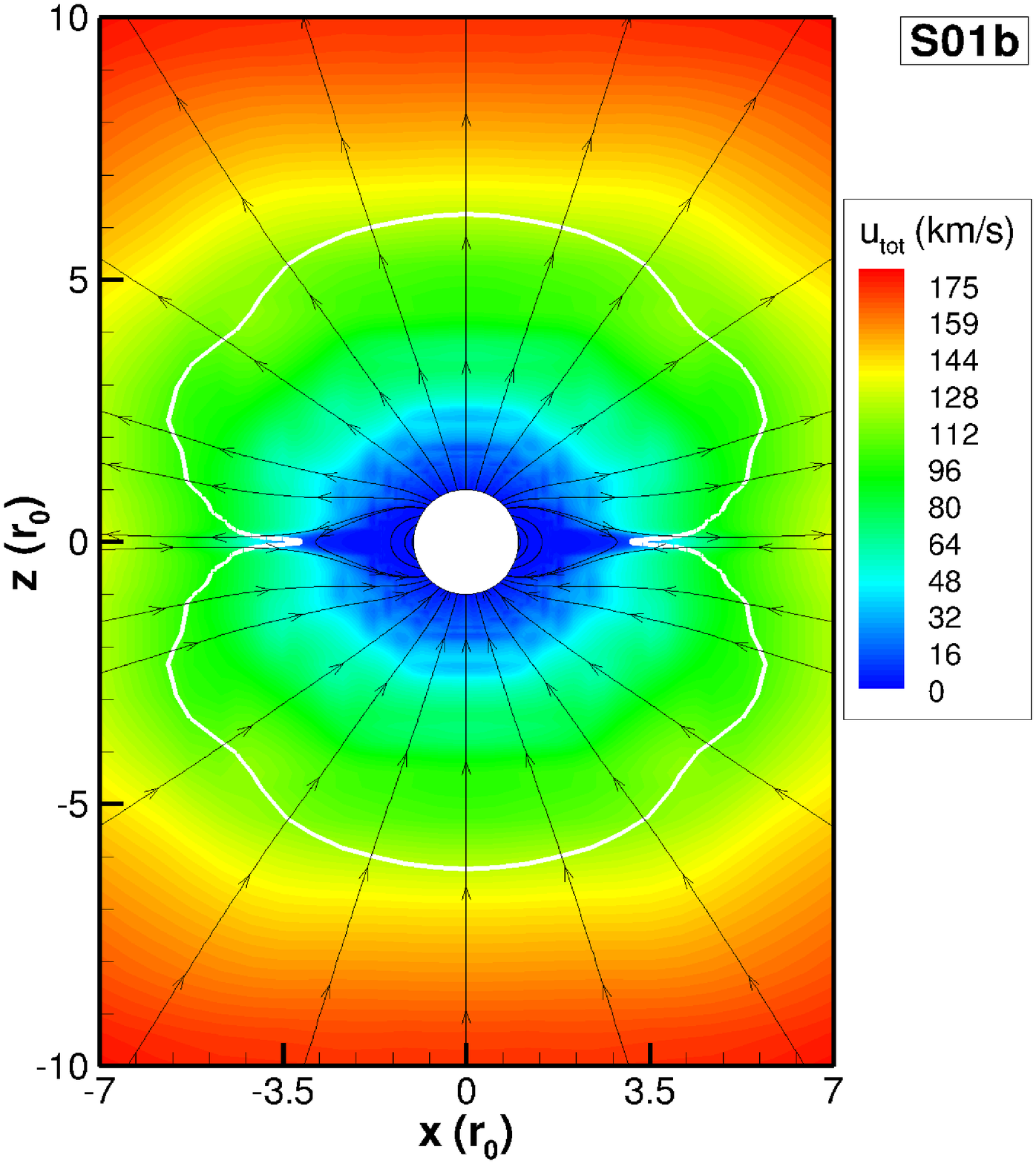} \\ %
\includegraphics[scale=0.26]{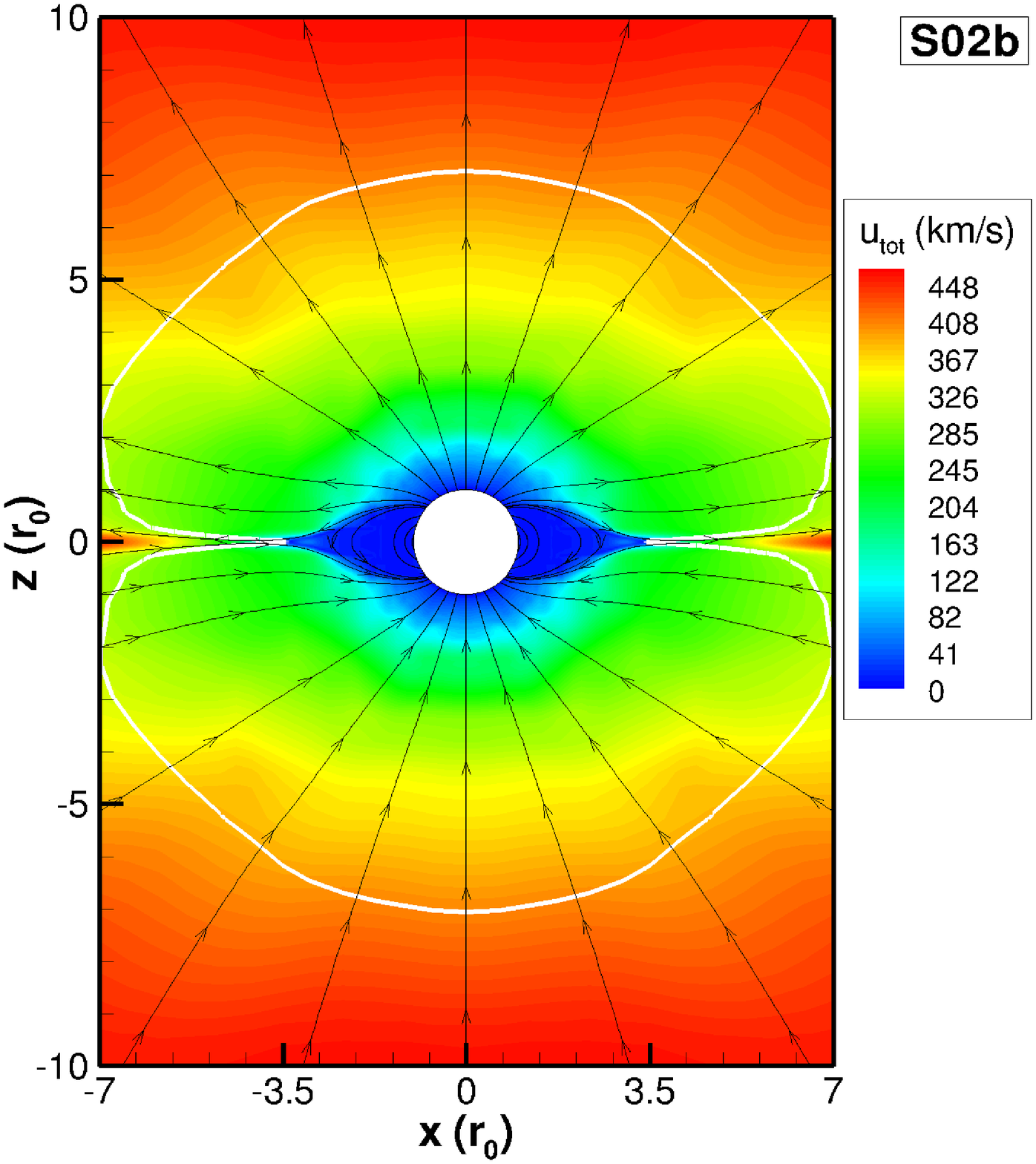}%
  \caption{Same as Fig.~\ref{ss.firstset.utot}, but for simulations S01b and S02b, where $\gamma=1.1$. \label{ss.thirdset}}
\end{figure}

We investigate now how the wind structure will change if a different $\gamma$ is considered. Considering the cases S01/S09 and S02/S08, where $\beta_0=1$, $1/25$, respectively, we performed simulations considering $\gamma=1.1$. This third set of simulations is presented in Table~\ref{table:1} and the meridional cuts of the steady-state configurations for S01b and S02b are shown in Fig.~\ref{ss.thirdset}. 

It is worth noting that by adopting a different $\gamma$, the acceleration mechanism of the wind changes, what changes the velocity profile of the wind and, consequently, the configuration of the magnetic field (compare, for instance, the steady-state configuration of S01 and S01b). This is expected because, the higher $\gamma$ is, the thermal acceleration mechanism is less effective (e.g., there is less turbulence or dissipation of waves).

Due to the scaling relations described in \S\ref{sec.comp}, simulations S01b and S09b, with $\beta_0=1$, present the same configuration of the steady-state wind. The same result is achieved for simulations S02b and S08b, with $\beta_0=1/25$.

\section{DISCUSSION AND CONCLUSION}
This is the first study to perform 3D MHD numerical simulations analyzing in detail winds of magnetized solar-like stars with different $\beta_0$. As initial condition, our model assumes a bipolar configuration for the magnetic field at the stellar surface. This distribution of magnetic field, although very simplified, is useful as a first step in understanding how the interaction between stellar magnetic field and the wind occur in a 3D self-consistent manner. 

In our simulations, we adopt a simpler treatment and parametrize the energy content of the wind in terms of $\gamma$. This is a first step towards more realistic simulations. It should be stressed that if one aims to describe the complex physics of stellar winds, a more complete description of the energy content of the wind has to be taken into account. In the solar corona, for instance, it is inferred that closer to the Sun, $\gamma\simeq 1$ \citep{1988JGR....9314269S}, and at $1$~AU, $\gamma \simeq 1.5$ \citep{1995JGR...100...13T,1995GeoRL..22.3301P}. Therefore, $\gamma$ is not expected to be constant throughout the stellar wind, as we assumed. As the aim of our study is to investigate the effects of the magnetic fields in the general properties of the stellar wind, the inclusion of MHD waves or variable $\gamma$ is postponed for future work.

In the first set of simulations where  a fixed heating parameter $\gamma=1.01$ was adopted, we showed that for a solar-like star, the increase in the magnetic field intensity creates faster winds, with general characteristics as the \citet{1971SoPh...18..258P} result for the Sun: a creation of a bi-modal stellar wind. The final configuration of the magnetic field consists of a zone of closed magnetic loops at low latitudes. On top of the closed loop region lies the neutral point, where the reconnection takes place and is the starting point of the current sheet that extends for larger distances along the equatorial plane. The zone of open field lines located at high latitudes of the star fills all the space outside the closed loop region and carries along a wind.

For a given $\beta_0$, the wind is more accelerated in the polar regions than in the equatorial region. This difference is generated by the latitude-dependent Lorentz force {which creates } a flux of matter directed to the equatorial plane. This generates a density enhancement at low latitudes providing a mass-loss rate per unit solid angle that is also latitude-dependent and increases at low latitudes. This meridional flux is intensified for higher magnetic fields. The increase in the magnetic field also led to hotter winds with higher mass-loss rates.

We showed that the $\beta$-parameter is a key parameter in the structure of the wind in the case $\gamma$ is the same for all simulations. As $\beta_0$ was increased back to $1$ by changing $\rho_0$ and maintaining $B_0=20$~G in the second set of simulations, the wind decelerated and cooled. Comparing the first and second sets of simulations, we conclude that winds with same $\beta_0$ have i) same velocity profile and ii) same magnetic field topology (the neutral point is located at the same position). However, mass-loss rates are different as $\dot m (\theta)$ is dependent on the choice of $\rho_0$.

By analyzing the effects of a different $\gamma$ in the simulations, we showed in the third set of simulations that the heating parameter $\gamma$ is, together with the plasma-$\beta$ parameter at the coronal base, an important parameter in the structure of the wind. However, the steady-state configuration of the wind in the third set is different from the first and second ones because the thermal force driving the wind changes when $\gamma$ changes. 

In the simulations we ran, we adopted $\gamma=1.01$ in the first and second sets, and $\gamma=1.1$ in the third set. Although $\gamma$ is necessary for the thermal acceleration of the wind, a wind with a low $\gamma$ implies a proportionally high Lorentz force, when compared to the thermal force. For this reason, the third set of simulations is more asymmetrical than the first set (compare S01 and S01b, S02 and S02b), since the Lorentz force is $\theta$-dependent. Another aspect is that in a purely HD coronal wind, the thermal force is the main responsible for the acceleration of the wind. This implies that by adopting a low $\gamma$, the wind has less internal energy available, being less thermally accelerated. In the cases S01 ($\gamma=1.01$) and S01b ($\gamma=1.1$), the same situation holds: case S01b is less accelerated than case S01. However, cases S02 ($\gamma=1.01$) and S02b ($\gamma=1.1$) show a different behavior. The present simulations indicate that a change in gamma from $1.01$ to $1.1$ decreases the terminal speed by $43\%$ in cases S01/S01b and increases $8\%$ in cases S02/S02b. This different behavior is because in case S02b, the Lorentz force provides a significant acceleration mechanism for the wind, i.e., the Lorentz force is proportionally more important than the thermal force. This illustrates the need for perform full 3D MHD simulations in order to assess the importance of $\beta$ and $\gamma$.

The effective temperature of solar-like stars is orders of magnitude lower than the coronal temperature adopted in our models. In the present work, we did not take into account what is heating the wind from photospheric temperatures up to coronal temperatures. Instead, we placed the inner boundary of our model at the coronal base. For a more realistic study, however, the energy equation should be solved starting from lower layers. Hence, the plasma-$\beta$ parameter at the coronal base would then be incorporated as resultant from the energy interactions happening in these lower layers rather than being a free parameter at the coronal base, as in our models. Models and 1D simulations that resolves the energy balance in the chromosphere-corona transition region have been calculated recently for the solar wind \citep{2005ApJ...632L..49S, 2007ApJS..171..520C} and stellar winds \citep{2007ApJ...659.1592S, 2008ApJ...689..316C}. If incorporated in 3D global simulations, these models can provide physical insights on the processes that accelerate and heat stellar winds.

The present work considered moderate $\beta_0$. In the presence of a weak magnetic field (high $\beta_0$), the wind is energetic enough to drag the field lines with it, leading to a configuration of radial magnetic field lines. Consequently, a spherical expansion of the corona is expected, similar to a purely HD case.

Here, we deal with a class of non-rotating stars. The validity of our results should be tested also in rotating objects, since a more generalized description of magnetized outflows from solar-like stars requires the inclusion of rotation. The detailed interaction including rotation requires detailed 3D MHD simulations. A future work will extend the present work to this direction.

\acknowledgments
AAV thanks the Brazilian agencies FAPESP and CAPES for financial supports under grants 04-13846-6 and BEX4686/06-3, respectively. AAV also thanks the warm reception at George Mason University where part of the work was performed. MO acknowledges the support by National Science Foundation CAREER Grant ATM-0747654, and FAPESP support (2007/58793-5) during her visits to Brazil. MO is also thankful for the hospitality of University of S\~ao Paulo. VJ-P thanks CNPq under grant (305905/2007-4). We would like to thank the staff at NASA Ames Research Center for the use of the Columbia supercomputer and the anonymous referees for comments and suggestions that helped improve the manuscript.


\end{document}